\documentclass[]{cbxarticle-ieee}

\usepackage{svg}
\usepackage{py}
\usepackage{multirow}
\usepackage{authblk}

\definecolor{dmlgreen}    {RGB}{51,  160,  44}
\definecolor{dmlblue}     {RGB}{31,  120, 180}
\definecolor{dmlred}      {RGB}{202,   0,  32}
\definecolor{mauve}       {rgb}{0.58,  0,0.82}

\lstdefinelanguage[RISC-V]{Assembler}
{
  alsoletter={.}, % allow dots in keywords
  alsodigit={0x}, % hex numbers are numbers too!
  morekeywords=[1]{ % instructions
    lb, lh, lw, lbu, lhu,
    sb, sh, sw,
    sll, slli, srl, srli, sra, srai,
    add, addi, sub, lui, auipc,
    xor, xori, or, ori, and, andi,
    slt, slti, sltu, sltiu,
    beq, bne, blt, bge, bltu, bgeu,
    j, jr, jal, jalr, ret,
    scall, break, nop,
    vadd.vv, vmsltu.vv, vpopc.m,
    vadd.vx, vmlst.vv, bnez,
    vsub.vv,
    mlxe.t, msxe.t, mszipk.tt, mszipv.tt, mssortk.tt, mssortv.tt,
    mmv.vi, mmv.vo
  },
  morekeywords=[2]{ % sections of our code and other directives
    .align, .ascii, .asciiz, .byte, .data, .double, .extern,
    .float, .globl, .half, .kdata, .ktext, .set, .space, .text, .word
  },
  morekeywords=[3]{ % registers
    zero, ra, sp, gp, tp, s0, fp,
    t0, t1, t2, t3, t4, t5, t6,
    s1, s2, s3, s4, s5, s6, s7, s8, s9, s10, s11,
    a0, a1, a2, a3, a4, a5, a6, a7,
    ft0, ft1, ft2, ft3, ft4, ft5, ft6, ft7,
    fs0, fs1, fs2, fs3, fs4, fs5, fs6, fs7, fs8, fs9, fs10, fs11,
    fa0, fa1, fa2, fa3, fa4, fa5, fa6, fa7,
    tr0, tr1, tr2, tr3, tr4, tr5, tr6, tr7, tr8,
    v0, v1, v2, v3, v4, v5, v6, v7, v8, v9,
    v10, v11, v12, v13, v14, v15, v16, v17, v18, v19,
    v20, v21, v22, v23, v24, v25, v26, v27, v28, v29,
    v30, v31
  },
  morecomment=[l]{;},   % mark ; as line comment start
  morecomment=[l]{\#},  % as well as # (even though it is unconventional)
  morestring=[b]",      % mark " as string start/end
  morestring=[b]'       % also mark ' as string start/end
}

\lstdefinestyle{riscv-asm-style}{
  language=[RISC-V]Assembler,                   % all code is RISC-V
  literate={ö}{{\"o}}1
           {ä}{{\"a}}1
           {ü}{{\"u}}1,
  basicstyle=\footnotesize\ttfamily,%
  breaklines=true,                              % break long lines
  keywordstyle=[1]\color{blue!80!black},        % instructions are blue
  keywordstyle=[2]\color{orange!80!black},      % sections/other directives are orange
  keywordstyle=[3]\color{red!50!black},         % registers are red
  stringstyle=\color{mauve},                    % strings are from the telekom
  commentstyle=\color{red}\ttfamily,%
  identifierstyle=\color{teal},                 % user declared addresses are teal
  tabsize=4,                                    % indent tabs with 4 spaces
  showstringspaces=false,                       % do not replace spaces with weird underlines
  numbers={left},%
  numbersep={5pt},%
  numberstyle={\tiny},%
  keepspaces={true},%
  upquote={true},%
  columns={fullflexible},%
  firstnumber=1
}

\lstdefinestyle{cpp-style}{%
  language=C++,%
  basicstyle=\footnotesize\ttfamily,%
  deletekeywords={if, else, for, static_cast},%
  morekeywords={parallel_for, parallel_invoke, parallel_reduce, uint8_t,
                uint_16_t, uint32_t, uint64_t, int8_t, int_16_t, int32_t,
                int64_t, Fault, Trace, InstRecord, ExecContext,
                RoccPacketPtr},%
  morekeywords=[2]{if, else, for, while, static_cast},%
  morekeywords=[3]{spawn, wait,  \%(op_src_decl)s, \%(op_rd)s,
                   \%(class_name)s, \%(op_wb)s, def},%
  keywordstyle=\color{NavyBlue}\ttfamily,%
  keywordstyle=[2]\color{Blue}\ttfamily,%
  keywordstyle=[3]\color{BurntOrange}\ttfamily,%
  stringstyle=\color{ForestGreen}\ttfamily,%
  commentstyle=\color{red}\ttfamily,%
  morecomment=[l][\color{Violet}]{\#},%
  numbers={left},%
  numbersep={5pt},%
  numberstyle={\tiny},%
  keepspaces={true},%
  upquote={true},%
  columns={fullflexible},%
  firstnumber=1
}

\lstdefinestyle{python-style}{%
  language=Python,%
  alsoletter={.},
  morekeywords=[2]{
    @pytest.mark.parametrize,
    @s.tick_fl,
    @s.tick_cl,
    @s.tick_rtl,
    @s.combinational,
    s.connect,
    s.connect_auto,
    s.connect_dict,
    Bits,
    Wire,
    InPort,
    OutPort,
    SimulationTool,
    TranslationTool,
    Model,
    UNDEFINED,
    integer,
    array,
    bit,
    reverse, sort, swap, compress, popc
  },
  morekeywords=[3]{
    SREG, VREG, TREG, ICREG, OCREG, MEM
  },
  basicstyle={\footnotesize\ttfamily},%
  keywordstyle=[1]\color{blue!80!black},
  keywordstyle=[2]\color{orange!80!black},
  keywordstyle=[3]\color{red!50!black},
  commentstyle=\color{red}\ttfamily,%
  numbers={left},%
  numbersep={5pt},%
  numberstyle={\tiny},%
  showstringspaces={false},%
  keepspaces={true},%
  upquote={true},%
  columns={fullflexible},
}%

\graphicspath{{../images/}{../prebuild/}}

\begin{document}

%-------------------------------------------------------------------------
% Front Matter
%-------------------------------------------------------------------------

\title
{%
  SparseZipper: Enhancing Matrix Extensions to \\Accelerate SpGEMM on CPUs
  %SparseZipper: Augmenting Matrix Extensions to \\Accelerate SpGEMM on CPUs
  %SparseZipper: Enhancing Matrix Extensions to \\Accelerate SpGEMM on CPUs
  %SparseZipper: An Augmentation of Matrix Extensions to Accelerate SpGEMM on CPUs
  % This sounds like we're not changing the existing matrix extensions
  %SparseZipper: Accelerating SpGEMM using Matrix Extensions on CPUs
  %Accelerating Merge-based Sparse Matrix Computations in Emerging Matrix Architectures
}

%\papernum{434}
%\confabbr{MICRO'24}

%\conference
%{%
%  Int'l Symp. on Microarchitecture (MICRO), 2024
%}

\author[1]{Tuan Ta}
\author[2]{Joshua Randall}
\author[1]{Christopher Batten}
\affil[1]{
  Cornell University, Ithaca, NY \authorcr {\{qtt2,cbatten\}@cornell.edu\vspace{1.5ex}}
}
\affil[2]{
  Arm Inc., Austin, TX \authorcr {Joshua.Randall@arm.com \vspace{-2ex}}
}

%\author
%{%
%  Tuan Ta, Joshua Randall, Christopher Batten
%}
%
%\affiliation
%{%
%  School of Electrical and Computer Engineering,
%  Cornell University, Ithaca, NY \\
%  cbatten@cornell.edu
%}

\maketitle

%-------------------------------------------------------------------------
% Body
%-------------------------------------------------------------------------

\renewcommand{\baselinestretch}{0.9}

\begin{abstract}
  The importance of general matrix multiplication (GEMM) is motivating
  new instruction set extensions for multiplying dense matrices in almost
  all contemporary ISAs, and these extensions are often implemented using
  high-performance systolic arrays. However, matrices in emerging
  workloads are not always dense, and sparse matrices where the vast
  majority of values are zeros are becoming more common. Existing matrix
  extensions and micro-architectures cannot efficiently process highly
  sparse matrices due to two reasons: (1)~wasted work when one or both
  input values are zero; and (2)~incompatibility with sparse matrix
  formats. This work proposes SparseZipper that minimally modifies
  existing matrix extensions and systolic-array-based micro-architectures
  specialized for dense-dense GEMM to accelerate sparse-sparse GEMM
  operating on highly sparse matrices with unstructured sparsity structures.
  Our performance evaluation shows SparseZipper achieves 5.98$\times$ and
  2.61$\times$ speedup over a scalar hash-based implementation of SpGEMM
  and a state-of-the-art vectorized SpGEMM version, respectively. Our
  component-level area evaluation shows SparseZipper increases the area
  of a baseline 16$\times$16 systolic array by only 12.7\% resulting in
  an area overhead for an entire system-on-chip of just a few percent.
\end{abstract}

%=========================================================================
% SparseZipper: Introduction
%=========================================================================

\section{Introduction}
\label{sec-spz-intro}

% *** Hardware trend: matrix extensions on CPUs ***
%   + Matrix extensions for CPUs are coming
%   + Why? --> accelerate dense matrix-matrix multiplications
%   + How? --> in the form of large 2D MAC array and work closely with existing vector architectures
%   + Benefits? --> being close to CPUs

General matrix multiply (GEMM) is a key building block in many different
domains including machine learning, graph analytics, and scientific
computing. Therefore, numerous domain-specific architectures have been
proposed to accelerate dense-dense GEMM (i.e., most values in both input
matrices are non-zeros) with various trade-offs in programmability,
performance, and energy
efficiency~\cite{jouppi-datacenter-isca2017,teich-google-tpu-v2-blog2018,chen-eyeriss-v2-jetcas2019,jouppi-google-tpu-v2-v3-cacm2020,choquette-tensor-core-nvidia-ieeemicro2021}.
In addition to coarse-grain accelerators, CPU vendors have recently
introduced matrix extensions (e.g., Intel's Advanced Matrix Extension
(AMX)~\cite{intel-amx-web,nassif-intel-sapphire-isscc2022,jeong-rasa-dac2021},
Arm's Scalable Matrix Extension (SME)~\cite{arm-sme-web}, RISC-V's matrix
extension proposal~\cite{riscv-mtx-ext-proposal-web}, and IBM's
Matrix-Multiply Assist (MMA)~\cite{ibm-mmx-assist-web}) to their ISAs for
dense-dense GEMM acceleration. Such matrix extensions attempt to strike a
balance between programmability and efficiency, and they are often
implemented using systolic-array-based
micro-architectures~\cite{intel-amx-web,nassif-intel-sapphire-isscc2022}.

% *** Software trend: Sparse computation ***
%   + Data are sparse and in many cases extremely sparse -> motivate the challenge
%   + Unstructured sparsity -> motivate the importance of a general solution
%   + Why dense hardware is no longer enough to do sparse?

However, matrices in workloads are not always dense. In fact, many recent
neural network
models~\cite{reddi-mlperf-isca2020,naumov-dnn-model-arxiv2019,han-deep-compress-arxiv2015,jouppi-datacenter-isca2017,wu-ml-facebook-hpca2019},
real-world graph
analytics~\cite{davis-graphblas-tmos2019,hoefler2011generic,shun-multicore-tc-2015},
and scientific
simulations~\cite{canning-sparse-sim-1996,galli-quantum-sim-1996} operate
on sparse matrices where the majority of values are zeros. In addition,
matrix densities (i.e., the percentage of non-zero values in a matrix)
vary dramatically across domains (e.g., from $10^{-6}\%$ density in
matrices representing social graphs to 50\% density in matrices used in
neural network models~\cite{hegde-extensor-micro2019}). Such low matrix
densities prevent computing GEMM for sparse matrices efficiently on CPUs
using the recently introduced matrix extensions since most
multiplications will involve at least one input value which is zero.
Moreover, sparse matrices are typically stored in compact formats with
metadata indicating positions of non-zero values for space efficiency, so
they are not directly compatible with existing built-in matrix engines
specialized for processing matrices stored in a dense format.

% *** Existing solutions (only closely related work) ***
%   + SparseCore - specialized ISA extension just for sparse computation
%   + VEGETA
%   + SparseTPU? Should we include this? It's not really an ISA extension

In addition to numerous domain-specific sparse-sparse GEMM (SpGEMM)
accelerators~\cite{qin-sigma-hpca2020, hegde-extensor-micro2019,
  zhang-sparch-hpca2020, srivastava-matraptor-micro2020,
  zhang-gamma-asplos2021}, previous work has proposed several ISA
extensions to accelerate sparse computations.
SparseCore~\cite{rao-sparsecore-asplos2022} is a stream-based ISA
extension designed specifically for sparse computations at the cost of
extra hardware for stream registers and stream processing units without
efficiently supporting dense-dense GEMM. VEGETA extends a matrix
extension to accelerate sparse-dense matrix-matrix multiplication (SpMM)
in addition to dense computations~\cite{jeong-vegeta-hpca2023}. However,
VEGETA is limited to SpMM and DNN-specific sparsity structures, so it is
not efficient when multiplying two highly sparse (i.e., less than 1\%
density) matrices with unstructured sparsity structures, which is critical in
various workload domains including graph analytics (e.g., multi-source
breadth-first search, peer pressure clustering, cycle detection, triangle
counting,
etc.)~\cite{dalberto-all-pair-spgemm-2007,shah-graph-spgemm-2007,rabin-maximum-matching-alg-1989,azad-triangle-count-2015},
hybrid linear solvers (e.g., Schur complement method and algebraic multi-grid
methods)~\cite{yamazaki-spgemm-schur-2010}, context-free grammar
parsing~\cite{penn-context-free-grammar-2006}, molecular dynamics
simulatio~\cite{itoh-order-n-spgemm-1995}, and interior point
methods~\cite{karypis-interior-point-alg-1994}.

% *** Our solution ***
% Augment matrix extensions to accelerate SpGEMM

In this work, we propose SparseZipper that minimally extends existing
matrix ISAs and systolic-array-based micro-architecture specialized for
dense-dense GEMM to accelerate SpGEMM operating on highly sparse matrices
with unstructured sparsity structures. SparseZipper targets a conventional
row-wise dataflow SpGEMM algorithm (i.e., Gustavson algorithm) with sparse
matrices represented in commonly used compressed sparse row/column
(CSR/CSC) formats. The abstraction and micro-architecture of SparseZipper
are specialized for accelerating the algorithm's main performance
bottleneck which involves merging multiple sparse vectors represented as
streams of indices (i.e., keys) and data (i.e., values).
By leveraging existing matrix registers for storing key-value streams and a
systolic array for merging multiple streams, SparseZipper incurs minimal area
overhead.
Our performance evaluation shows SparseZipper achieves 5.98$\times$ and
2.61$\times$ speedup over a scalar hash-based implementation of SpGEMM and a
state-of-the-art vectorized SpGEMM version, respectively.
Our component-level area evaluation shows SparseZipper increases the area of a
baseline 16$\times$16 systolic array by only 12.7\%.
This overhead would be much lower when considering an entire processor and its
caches.

%SparseZipper targets the key bottleneck, which is merging partial sparse
%vectors, in a conventional SpGEMM algorithm for data-parallel
%architectures~\cite{li-merge-spmspv-vectorarch-2018,li-spgemm-vector-arch-mchpc2019,fevre-spgemm-rvv-arxiv2023,winter-adaptive-spgemm-gpu-ppopp2019,liu-efficient-spgemm-gpu-ipdps2014,dalton-optimizing-spgemm-gpu-ipdps2015}.
%Each partial sparse vector is considered as a stream of keys (i.e.,
%representing row/column indices of non-zeros in a matrix) and corresponding
%non-zero values.
%At the core of SparseZipper is its ability to efficiently merge such streams in
%parallel by leveraging in-place matrix registers to store parts of concurrent
%streams and built-in systolic array to merge those streams together.
%In order to facilitate that merge operation, we propose a minimal set of
%additional architectural states to keep track of active streams and matrix
%instructions to move streams between matrix registers and memory.
%Our performance evaluations show SparseZipper achieves 5.98$\times$ and
%2.61$\times$ speedup over a scalar hash-based implementation of SpGEMM and a
%vectorized SpGEMM version respectively.
%Our post-synthesis area evaluation shows SparseZipper incurs less than XX\%
%area overhead compared to the baseline matrix engine designed for dense GEMM.

% *** Our contributions ***
\BF{Contributions --} Our key contributions include: (1)~a SparseZipper
ISA extension that enhances an existing matrix ISA to efficiently support
merging multiple key-value streams, the main performance bottleneck in
the conventional row-wise dataflow SpGEMM algorithm; (2)~a minimal set of
micro-architectural changes to a systolic array to support the new
SparseZipper instructions; and (3)~a detailed cycle-level evaluation
demonstrating the performance benefits of SparseZipper and a first-order
area evaluation demonstrating the minimal additional hardware needed for
SparseZipper.

\section{Background}
\label{sec-spz-background}

%This section provides background on recent matrix ISA extensions and three
%common SpGEMM dataflows: inner-product, outer-product, and row-wise-product.
This section provides background on recent matrix ISA extensions and common
SpGEMM algorithms.

\subsection{Matrix ISA Extensions for Dense GEMM}

The importance of GEMM has led to an emergence of matrix extensions in
contemporary ISAs.
Arm recently released its SME extension that introduces a new instruction
performing an outer product of two vectors and accumulating its results into a
new two-dimensional accumulator register~\cite{arm-sme-web}.
IBM took a similar approach in its MMA extension for the Power
ISA~\cite{ibm-mmx-assist-web}.
Intel introduced a new AMX extension that adds several matrix registers called
tile registers and a new matrix-matrix multiply instruction on two tile
registers~\cite{intel-amx-web,nassif-intel-sapphire-isscc2022}.
The RISC-V community recently proposed a matrix extension that is similar to
Intel AMX's approach~\cite{riscv-mtx-ext-proposal-web}.
One common micro-architecture for accelerating dense-dense GEMM is a systolic
array, a grid of multiply-add processing elements (PEs) connected in a mesh
network~\cite{jouppi-datacenter-isca2017,jeong-rasa-dac2021,nassif-intel-sapphire-isscc2022}.
A systolic array can support either input-, weight-, or output-stationary
dataflows, depending on which input or output matrix stays inside the array
throughout the computation.

\subsection{SpGEMM Dataflows}

The inner-product dataflow computes each element in the output matrix by
performing a dot product between a row in the first input matrix and a
corresponding column in the second input matrix.
Multiple dot product operations for different output elements can happen in
parallel.
For highly sparse matrices, one major downside of this dataflow is that a dot
product of two highly sparse vectors is likely to produce a zero, which is
wasted.

The outer-product dataflow performs an outer product between a column in the
first input matrix and a corresponding row in the second input matrix to
produce a partial output matrix.
Multiple partial output matrices are then merged into a single matrix.
This dataflow avoids the wasted computation incurred in the inner-product
dataflow at the cost of highly complex merging operation of multiple matrices
and potentially significant memory space for storing partial matrices.

The row-wise-product dataflow (Gustavson algorithm) computes each row
of an output matrix by multiplying a row in the first input matrix with the
entire second input matrix (vector-matrix multiplication).
Similar to the outer-product dataflow, this row-wise-product dataflow is
work-efficient for highly sparse matrices since it processes only non-zero
input elements that contribute to non-zero output elements.
The vector-matrix multiplication involves merging multiple sparse vectors into
a single vector, which is less complex than merging multiple sparse matrices as
in the outer-product dataflow.
In addition, unlike both inner-product and outer-product dataflows, the
row-wise-product dataflow does not require two input matrices to be stored in
two different formats: CSR and CSC.
All input and output matrices can be consistently stored in CSR, so there is no
need for converting between different sparse matrix formats.
In this work, we target the row-wise-product dataflow.

%\input{sec-background-mtx-arch}
%\input{sec-background-spgemm}
%=========================================================================
% SparseZipper: ISA extension
%=========================================================================

\section{S\lowercase{parse}Z\lowercase{ipper} Instruction Set Extension}
\label{sec-spz-isa}

In this section, we first describe a merge-based implementation of the
row-wise-product SpGEMM algorithm to motivate key designs in our SparseZipper
ISA extension.
We then present details of SparseZipper abstraction.

\begin{figure}[tp]
  \centering
  \includegraphics[width=\cw]{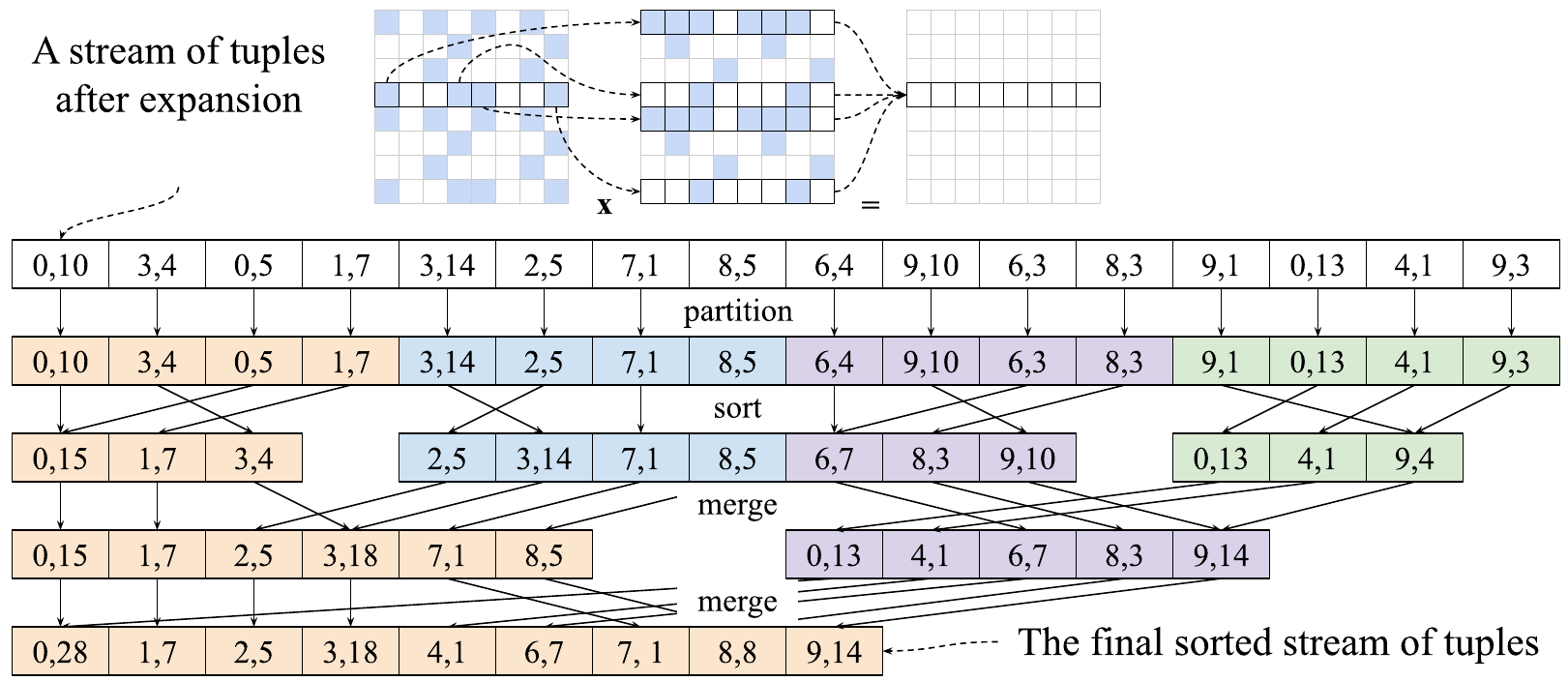}
  \caption{
    Multiple Steps to Compute One Output Row --
    Each tuple includes a column index (key) and a value.
  }
  \label{fig-spz-stream-merge-tree}
  \vspace{-0.2cm}
\end{figure}

\begin{figure}[!t]
  \centering
  \includegraphics[width=\cw]{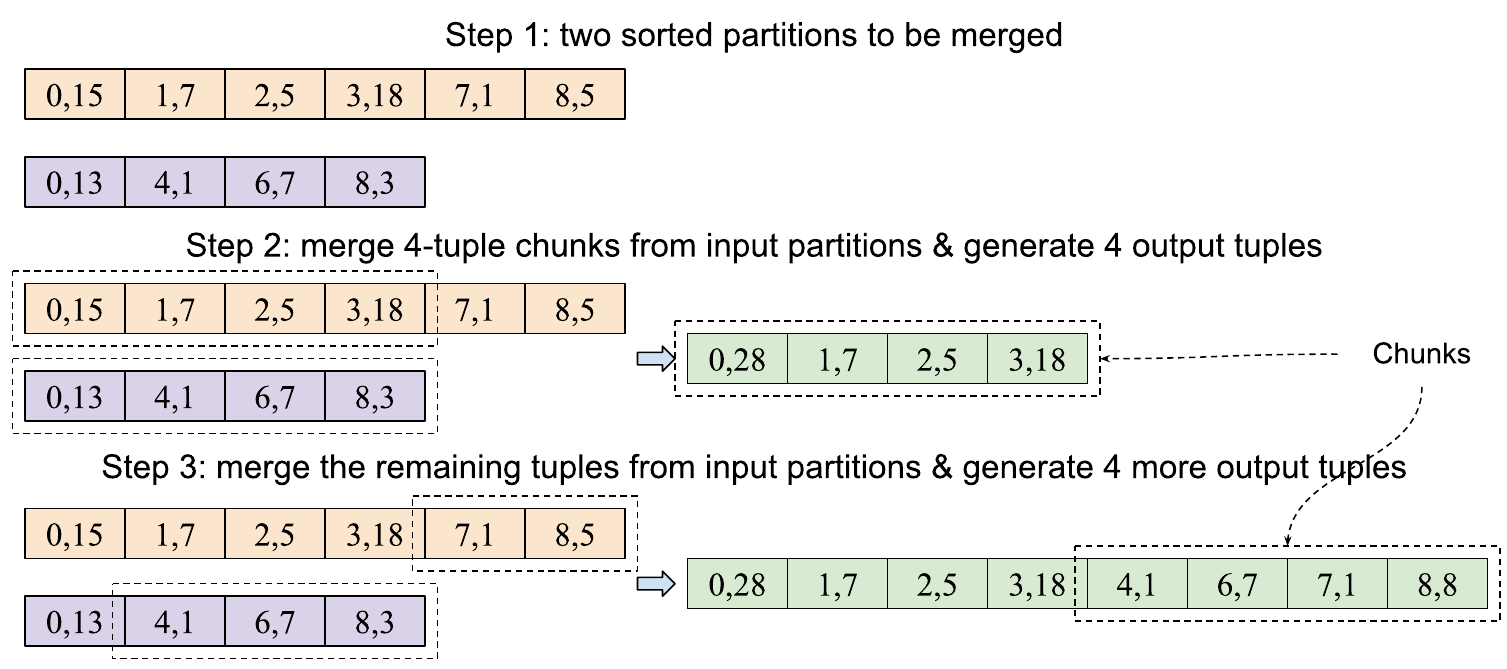}
  \caption{
    Merging Two Sorted Key-Value Partitions in Chunks -- e.g., merging the last two key-value partitions in Figure~\ref{fig-spz-stream-merge-tree}.
  }
  \label{fig-spz-stream-merging-steps}
  \vspace{-0.2cm}
\end{figure}

%----------
%----------
% Motivate the proposed instruction by showing how key-value streams are merged
% together in a merge tree (slightly modified merge sort algorithm)
\subsection{Merge-Based Row-Wise-Product SpGEMM}
\label{sec-spz-isa-merge-spgemm}

Figure~\ref{fig-spz-stream-merge-tree} shows an example of the row-wise-product
dataflow in multiplying two sparse matrices.
Partial results for an \IT{i-th} row in the output matrix are generated by
multiplying each non-zero element \TT{A[i][j]} in the first matrix with all
non-zero elements \TT{B[j][k]} in a \IT{j-th} row of the second matrix.
After generating partial results, for each row of the output matrix, we get
a list of tuples, each consisting of a column index (key) and a value.
To generate an output row, this list, which is called a key-value stream, is
then sorted by keys, and tuples with duplicate keys are accumulated.
The final sorted stream of unique key-value tuples represents non-zeros in the
\IT{i-th} row of the output matrix.

Merging partial tuples into the final list can be done in multiple steps.
The expanded list is split into equally sized partitions.
Tuples in each partition are then sorted by their keys.
Finally, adjacent partitions are merged together in multiple reduction steps to
form a final sorted streams of tuples, as shown in
Figure~\ref{fig-spz-stream-merge-tree}.
This merging procedure is similar to the conventional merge sort algorithm
except that tuples with duplicate keys are accumulated.

In order to merge two long sorted partitions of key-value tuples, we break them
in chunks fitting in registers (e.g., vector registers) and repeatedly merge
two N-element chunks (i.e., one from each tuple) at a time as shown in
Figure~\ref{fig-spz-stream-merging-steps}.
It is important to note that we may not be able to move all N tuples from each
partition in one step.
For example, in the first two chunks (in
Figure~\ref{fig-spz-stream-merging-steps}), three tuples (4, 1), (6, 7), and
(8, 3) from the second partition cannot be moved to the output partition since
their keys are greater than every key from the current chunk in the first
partition.
Instead, those tuples need to be merged in the next step.
Therefore, the number of tuples that we can advance at the end of a step in
each partition is data-dependent.

%----------
%----------
\subsection{Architectural States}
\label{sec-spz-isa-arch-states}
\begin{figure}[tp]
  \centering
  \includegraphics[width=\cw]{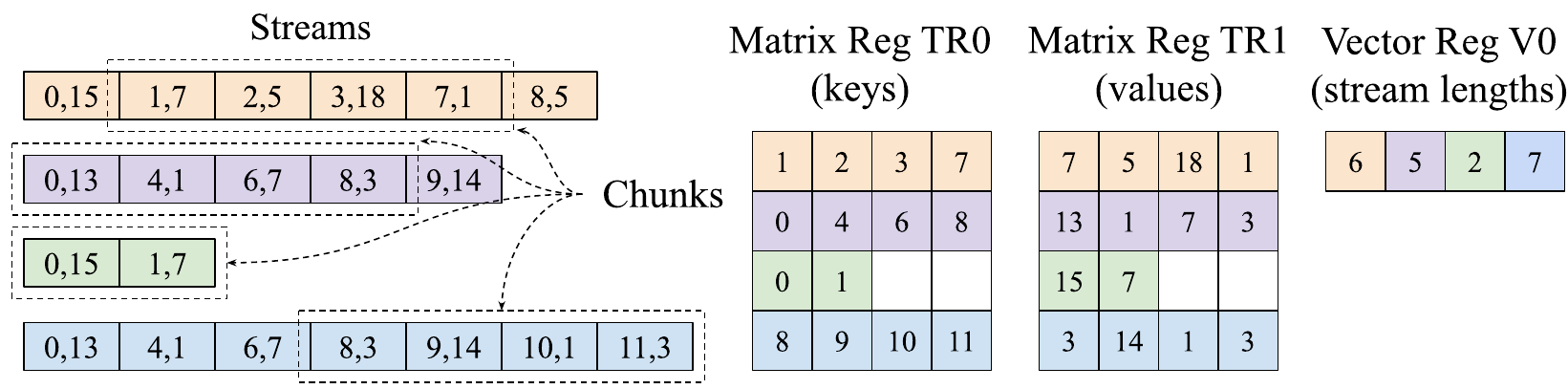}
  \caption{
    Mapping Between Key-Value Streams and Matrix Registers --
    Only chunks of key-value tuples with dashed borders are
    held in the two matrix registers.
  }
  \label{fig-spz-stream-mtx-reg-mapping}
  \vspace{-0.2cm}
\end{figure}

SparseZipper leverages both vector and matrix registers in the base vector and
matrix ISAs to store key-value streams and their metadata.
Without the loss of generality, in this work, we use the RISC-V vector
extension~\cite{riscv-vector-ext} as the base vector ISA and a baseline matrix
ISA inspired by Intel AMX~\cite{intel-amx-web} and the RISC-V matrix extension
proposal~\cite{riscv-mtx-ext-proposal-web}.

\BF{Matrix registers --}
The base matrix ISA supports eight general-purpose two-dimensional matrix
registers (tile registers) named from \TT{TR0} to \TT{TR7}.
The length in bits for a row in a matrix register is the same as the number of
bits in a single vector register (\TT{VLEN}) as defined in the RISC-V vector
extension.
In this work, we limit the size in bits of each element (i.e., \TT{ELEN}) in a
vector register to 32 bits to simplify our description of SparseZipper.
A complete matrix instruction set may support other element bit widths such as
16 and 64 bits.
There are $R = VLEN / ELEN$ elements in a row of a matrix register.
We assume that each matrix register has the same number of rows as the number
of elements in a row.
%Therefore, the total number of elements in a matrix register is $R^2$.

\BF{Mapping streams to matrix and vector registers --}
SparseZipper enables processing multiple key-value streams by mapping them to
different rows of a matrix register.
Figure~\ref{fig-spz-stream-mtx-reg-mapping} shows an example of this
stream-register mapping.
Matrix registers \TT{TR0} and \TT{TR1} store keys and values respectively.
Each row of a matrix register is mapped to a stream, and it stores a chunk of
keys or values in the stream.
SparseZipper uses existing vector registers to store metadata about a current
set of streams (e.g., vector register \TT{V0} in
Figure~\ref{fig-spz-stream-mtx-reg-mapping} stores the number of tuples in each
stream).

\BF{Special-purpose counter vector registers --}
SparseZipper introduces a set of four light-weight special-purpose counter
vector registers.
\TT{IC0} and \TT{IC1} are used for counting processed elements per input matrix
row.
\TT{OC0} and \TT{OC1} are used for counting output elements per output matrix
row.
More details regarding the use of those registers are specified in the
following instruction set specification.
Since each counter in a counter vector register counts up to the max number of
elements (i.e., \TT{R}) in a row of a matrix register, each counter is
$\log_2R$-bit wide.  Therefore, each counter vector register has $R \times
\log_2 R$ bits in total.

%----------
%----------
\subsection{Instruction Set Specification}
\begin{table*}[tp]
  \centering
  \cbxsetfontsize{8pt}
  \tabcolsep 3pt
  \caption{List of SparseZipper Instructions}
  \begin{tabular}{ll}
  \toprule
  \multicolumn{1}{c}{\textbf{Instructions}} & \multicolumn{1}{c}{\textbf{Description}}  \\
  \midrule
  %  \multicolumn{2}{l}{\textbf{Matrix Instructions for Dense GEMM (base matrix ISA)}}                                       \\
  %  \TT{mmult.tt td1, ts2, ts3}         & Multiply matrices in \TT{ts2} and \TT{ts3} and accumulate results into \TT{td1}   \\
  %  \TT{mlse.t td1, 0(rs1), rs2, vs3}   & Load matrix data into \TT{td1} using a constant stride \TT{rs2}                   \\
  %  \TT{msse.t ts1, 0(rs1), rs2, vs3}   & Store matrix data from \TT{ts1} using a constant stride \TT{rs2}                  \\
  %\midrule
  %  \multicolumn{2}{l}{\textbf{Matrix Instructions for Sparse GEMM (SparseZipper ISA extension)}}                 \\
    \TT{mlxe.t td1, 0(rs1), vs2, vs3}   & Load data into \TT{td1} using indices in \TT{vs2}; \TT{rs1} is the base address; \TT{vs3} are stream lengths.                      \\
    \TT{msxe.t ts1, 0(rs1), vs2, vs3}   & Store data from \TT{ts1} using indices in \TT{vs2}; \TT{rs1} is the base address; \TT{vs3} are stream lengths.                     \\
    \TT{mssortk.tt td1, td2, vs1, vs2}  & Sort keys in \TT{td1} and \TT{td2}; \TT{vs1} and \TT{vs2} are input lengths.                                                       \\
    \TT{mssortv.tt td1, td2, vs1, vs2}  & Shuffle \& accumulate values in \TT{td1} and \TT{td2} based on last key sorting results; \TT{vs1} and \TT{vs2} are input lengths.  \\
    \TT{mszipk.tt td1, td2, vs1, vs2}   & Merge keys in \TT{td1} and \TT{td2}; \TT{vs1} and \TT{vs2} are input lengths.                                                      \\
    \TT{mszipv.tt td1, td2, vs1, vs2}   & Shuffle \& accumulate values in \TT{td1} and \TT{td2} based on last key merging results; \TT{vs1} and \TT{vs2} are input lengths.  \\
    \TT{mmv.vi vd, cimm}                & Move values from an input counter vector \TT{IC[cimm]} to \TT{vd}       \\
    \TT{mmv.vo vd, cimm}                & Move values from an output counter vector \TT{OC[cimm]} to \TT{vd}      \\
  \bottomrule
  \end{tabular}
  \label{tab-spz-mtx-insts}
  \vspace{-0.3cm}
\end{table*}

Table~\ref{tab-spz-mtx-insts} shows a list of SparseZipper instructions
including (1) indexed matrix load and store, (2) stream sorting, (3) stream
merging, and (4) counter vector move instructions.

\BF{Indexed matrix load and store instructions --}
SparseZipper introduces two memory instructions: \TT{mlxe.t} and \TT{msxe.t} to
move key-value chunks from multiple streams between matrix registers and
memory.
Multiple key-value streams may have different lengths, and their chunks are
located at arbitrary locations.
Therefore, in addition to the base address (\TT{rs1}) and matrix register
(\TT{td1}), \TT{mlxe.t} and \TT{msxe.t} take two vector operands: \TT{vs2}
specifying memory locations (i.e., byte offsets to a base address) and \TT{vs3}
holding stream lengths.

\BF{Stream sorting instructions --}
SparseZipper introduces two instructions called \TT{mssortk.tt} and
\TT{mssortv.tt} to sort multiple chunks of key-value tuples by keys.
The two instructions work together by first sorting keys and then shuffling
values based on the key reordering.
Duplicate keys are combined, and their corresponding values are accumulated.
In order to transfer the key reordering information between \TT{mssortk.tt} and
\TT{mssortv.tt} instructions, SparseZipper adds an abstract special-purpose
architectural state that captures how input keys are reordered per key-value
chunk.
This state is intentionally left abstract in the ISA specification so that a
micro-architecture can freely choose how to implement it.
Section~\ref{sec-spz-uarch} later discusses an implementation of this state
using a systolic array.
Since an output chunk may be shorter than its input chunk (i.e., due to
duplicate keys), \TT{mssortk.tt} updates the special-purpose output counter
vector registers (\TT{OC0} and \TT{OC1}) with the lengths of output chunks.

\BF{Stream merging instructions --}
SparseZipper provides two instructions called \TT{mszipk.tt} and \TT{mszipv.tt}
to merge sorted key-value partitions of a stream.
Similar to \TT{mssortk.tt} and \TT{mssortv.tt}, the two stream merging
instructions work together by first merging keys and then shuffling values.
Duplicate keys are combined, and their corresponding values are accumulated.
The key reordering is also captured by an abstract special-purpose
architectural state that is produced by \TT{mszipk.tt} and then used by
\TT{mszipv.tt} to shuffle and accumulate values.
Instruction \TT{mszipk.tt} updates input counter vector registers (\TT{IC0} and
\TT{IC1}) with the number of tuples that have been merged per input partition.
The output counter vector registers (\TT{OC0} and \TT{OC1}) are updated with
the number of elements per merged output partition.

\BF{Counter vector move instructions --}
SparseZipper provides two move instructions \TT{mmv.vi} and \TT{mmv.vo} that
copy values from special-purpose input and output counter vector registers
respectively into general-purpose vector registers.
These values are typically used to update pointers and stream lengths through
vector instructions.

%----------
%----------
\subsection{Code Examples}

\lstset{style=riscv-asm-style}

\begin{figure*}[tp]
  \centering

  %---------------------------------------------------------------------------
  % mssort example
  %---------------------------------------------------------------------------

  \begin{subfigure}[t]{\cw}
    \centering
    \includegraphics[width=\cw]{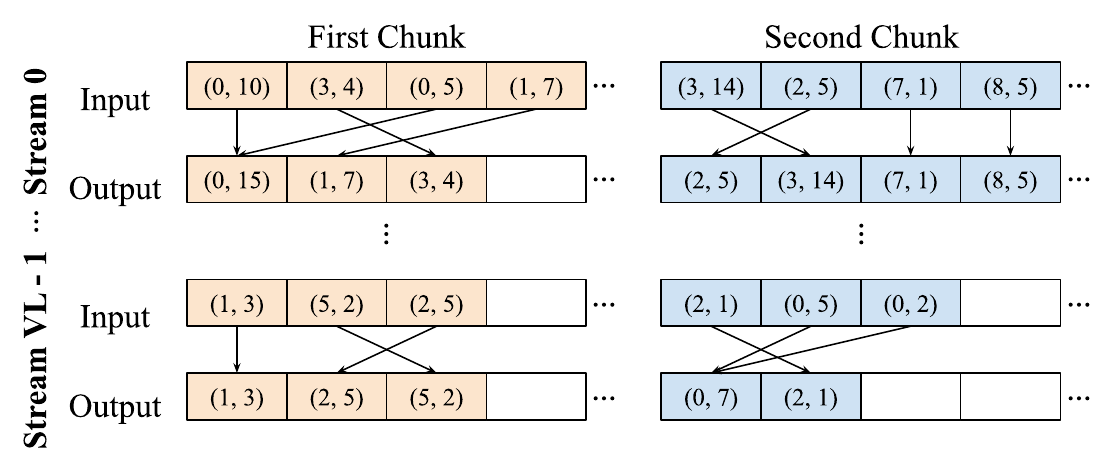}
    %\vspace{-0.8cm}
\begin{lstlisting}[xleftmargin=0.025\tw]
# a0, a1: base addresses of input key & value arrays
# a2, a3: base addresses of output key & value arrays
# v0, v1: lengths of the 1st & 2nd input chunks
# v2, v3: pointers to the 1st & 2nd input chunks
# v4, v5: lengths of the 1st & 2nd output chunks
# v6, v7: pointers to the 1st & 2nd output chunks
# load keys & values from both input chunks
mlxe.t tr0, 0(a0), v2, v0
mlxe.t tr1, 0(a1), v2, v0
mlxe.t tr2, 0(a0), v3, v1
mlxe.t tr3, 0(a1), v3, v1
# sort keys & values from both input chunks
mssortk.tt tr0, tr2, v0, v1
mssortv.tt tr1, tr3, v0, v1
# get lengths of output chunks
mmv.vo v4, 0x0
mmv.vo v5, 0x1
# store sorted keys & values to both output chunks
msxe.t tr0, 0(a2), v6, v4
msxe.t tr1, 0(a3), v6, v4
msxe.t tr2, 0(a2), v7, v5
msxe.t tr3, 0(a3), v7, v5
\end{lstlisting}
    \vspace{-0.4cm}
    \caption{Sorting Chunks of Keys and Values}
    \label{fig-spz-mssort-code}
  \end{subfigure}
  \hfill
  %---------------------------------------------------------------------------
  % mszip example
  %---------------------------------------------------------------------------
  \begin{subfigure}[t]{\cw}
    \centering
    \includegraphics[width=\cw]{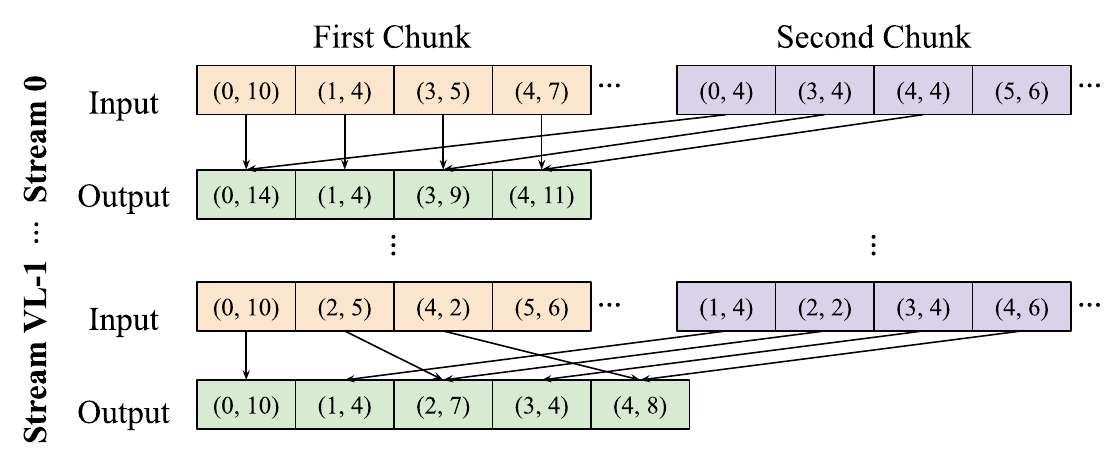}
    %\vspace{-0.8cm}
\begin{lstlisting}[xleftmargin=0.025\tw]
# a0, a1: base addresses of input key & value arrays
# a2, a3: base addresses of output key & value arrays
# v0, v1: lengths of the 1st and 2nd input chunks
# v2, v3: pointers to the 1st and 2nd input chunks
# v4: lengths of the output chunks
# v5: pointers to the output chunks
# load keys & values from both input chunks
mlxe.t tr0, 0(a0), v2, v0
mlxe.t tr1, 0(a1), v2, v0
mlxe.t tr2, 0(a0), v3, v1
mlxe.t tr3, 0(a1), v3, v1
# merge keys & values from both input chunks
mszipk.tt tr0, tr2, v0, v1
mszipv.tt tr1, tr3, v0, v1
# get the number of merged elements from input chunks
mmv.vi v6, 0x0
mmv.vi v7, 0x1
# get the number of elements added to output chunks
mmv.vo v8, 0x0
mmv.vo v9, 0x1
# store merged keys & values to output chunks
msxe.t tr0, 0(a2), v5, v8
msxe.t tr1, 0(a3), v5, v8
vadd.vv v5, v5, v8  # bump output pointers
msxe.t tr2, 0(a2), v5, v9
msxe.t tr3, 0(a3), v5, v9
vadd.vv v5, v5, v9  # bump output pointers
\end{lstlisting}
    \vspace{-0.4cm}
    \caption{Merging Chunks of Keys and Values}
    \label{fig-spz-mszip-code}
  \end{subfigure}
  \caption{
    Examples of Using SparseZipper Instructions to Sort and Merge Key-Value Streams --
    \TT{a\{0..3\}}~=~scalar registers;
    \TT{v\{0..9\}}~=~vector registers;
    \TT{tr\{0..3\}}~=~matrix registers.
  }
  \label{fig-spz-code}
  \vspace{-0.3cm}
\end{figure*}

%\input{fig-mssort-code}
%\input{fig-mszip-code}

%In this section, we show how to use the new SparseZipper instructions to sort
%and merge key-value chunks from multiple streams.

\BF{Sorting key-value chunks --}
Figure~\ref{fig-spz-mssort-code} shows a \mbox{RISC-V} assembly code snippet of
sorting key-value chunks from VLEN number of streams.
Keys and values of current chunks are loaded into matrix registers using
\TT{mlxe.tt} in lines 8-11.
Register \TT{tr0} and \TT{tr2} hold input keys while register \TT{tr1} and
\TT{tr3} store corresponding input values.
In line 13, \TT{mssortk.tt} sorts per-chunk keys in an ascending order and
writes the sorted keys in the same matrix registers (\TT{tr0} and \TT{tr2}).
In line 14, \TT{mssortv.tt} shuffles and accumulates values based on the
reordering of keys.
Lines 16-17 move lengths of output chunks from the special-purpose output
counter vectors into general-purpose vector registers for updating the lengths
of output streams.
Finally, keys and values in output chunks are written back to memory in lines
19-22.

\BF{Merging key-value chunks --}
Figure~\ref{fig-spz-mszip-code} shows a RISC-V assembly code snippet of merging
key-value chunks from adjacent partitions across VLEN number of streams.
Similar to the sorting code, keys and values of current chunks are loaded into
matrix registers using \TT{mlxe.tt} in lines 8-11, and register \TT{tr0} and
\TT{tr2} hold input keys while register \TT{tr1} and \TT{tr3} store
corresponding input values.
In lines 13-14, \TT{mszipk.tt} merges pairs of key-value chunks across VLEN
streams, and \TT{mszipv.tt} shuffles corresponding values based on the key
reordering.
Duplicate keys are combined, and their values are accumulated.
Merged and sorted output keys are stored in the same \TT{tr0} and then \TT{tr1}
in an ascending order.
Lines 16-17 move per-input-chunk numbers of merged keys from special-purpose
counter registers \TT{IC0} and \TT{IC1} into general-purpose vector registers
for updating input pointers.
Lines 19-20 extract lengths of output chunks from counter vector registers
\TT{OC0} and \TT{OC1}.
In lines 22-27, merged keys and values are written back to memory per output
streams using \TT{msxe.t}.

%=========================================================================
% SparseZipper: Micro-architecture
%=========================================================================

\section{S\lowercase{parse}Z\lowercase{ipper} Micro-Architecture}
\label{sec-spz-uarch}

This section describes SparseZipper micro-architecture that extends a baseline
systolic array specialized for dense-dense GEMM to support the proposed
instructions for sparse-sparse GEMM presented in Section~\ref{sec-spz-isa}.
Each sorting/zipping instruction is decomposed into micro-operations.
Each input matrix row corresponding to a data stream is processed in one
micro-operation going through the systolic array in two passes: (1)
sorting/zipping and (2) compressing.  The execution of one micro-operation is
explained in Section~\ref{sec-spz-uarch-mssort} and~\ref{sec-spz-uarch-mszip}.
In Section~\ref{sec-spz-uarch-multi-rows}, we then explain how pipelining
happens across micro-operations of a single instruction and multiple
instructions.
Finally, in Section~\ref{sec-spz-uarch-ext}, we discuss hardware changes needed
to support SparseZipper.

\begin{figure*}[tp]
  \centering
  \begin{subfigure}{\textwidth}
      \includegraphics[width=\textwidth]{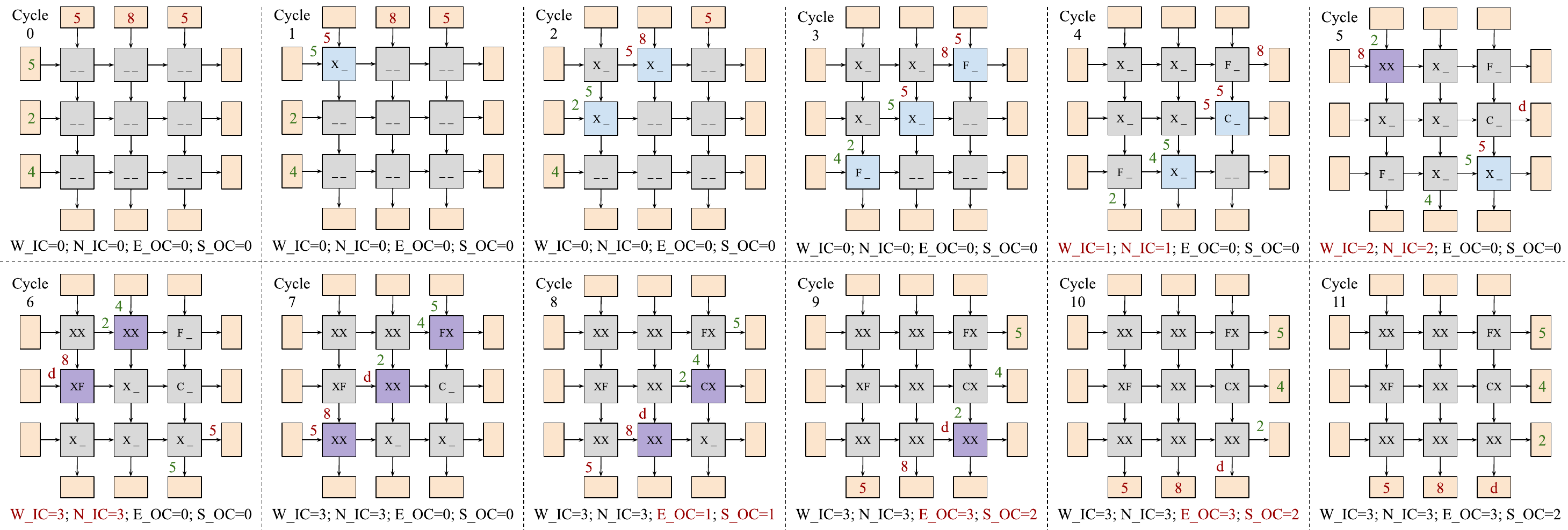}
      \caption{Systolic Execution of \TT{mssortk} Instruction.}
      \label{fig:mssortk-systolic-exec}
  \end{subfigure}
  \hfill
  \vspace{0.01cm}
  \begin{subfigure}{\textwidth}
      \includegraphics[width=\textwidth]{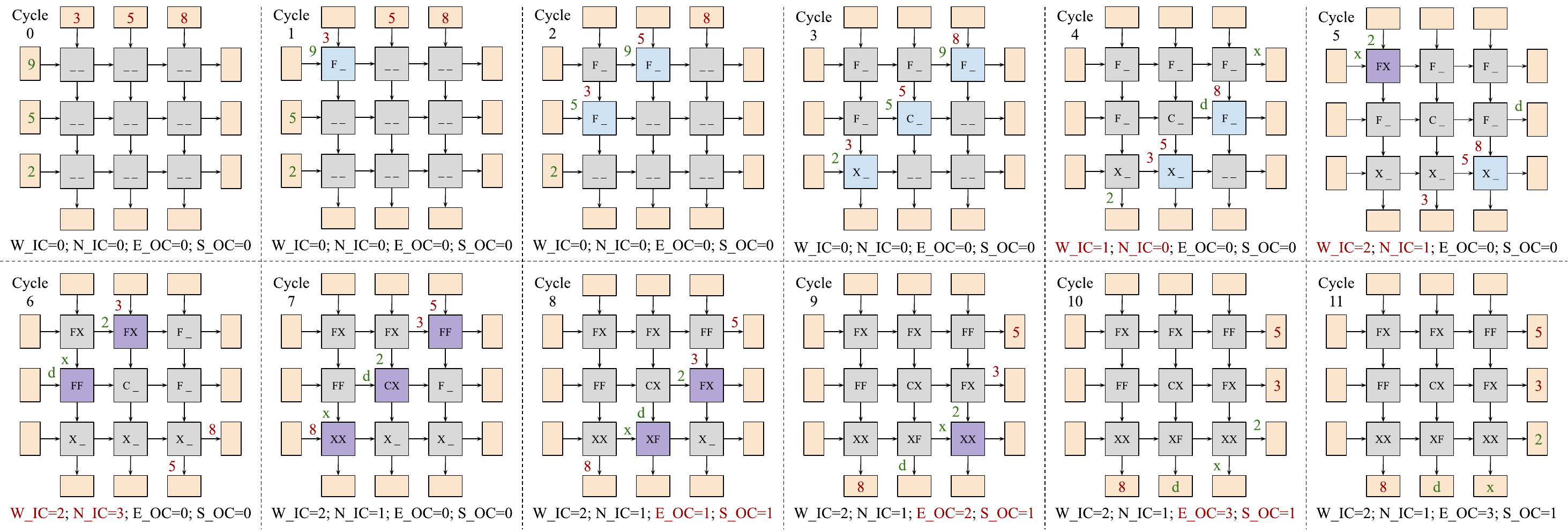}
      \caption{Systolic Execution of \TT{mszipk} Instruction.}
      \label{fig:mszipk-systolic-exec}
  \end{subfigure}
  \caption[Cycle-by-Cycle Systolic Execution of \TT{mssortk} in a 3$\times$3 Systolic Array for Two Unsorted Lists of Keys]{
    Cycle-by-Cycle Systolic Execution of \TT{mssortk} in a 3$\times$3 Systolic Array for Two Unsorted Lists of Keys --
    PE states: F~=~forward, X~=~switch, C~=~combine;
    W\_IC~=~west input counter;
    N\_IC~=~north input counter;
    E\_OC~=~east output counter;
    S\_OC~=~south output counter;
    d~=~duplicate key that is excluded;
    x~=~unmergeable key;
    Counters in red indicate they are being updated.
    PEs in gray are inactive.
    PEs in blue are merging keys.
    PEs in purple are compressing valid output keys.
    Keys in red come from the north input.
    Keys in green come from the west input.
    Keys in west and east sides are ordered from bottom to top.
    Keys in north and south sides are ordered from left to right.
  }
  \label{fig:systolic-exec}
  \vspace{-0.1cm}
\end{figure*}

%\lstset{style=riscv-asm-style}
\begin{figure*}[tp]
  \centering
  \includegraphics[width=\tw]{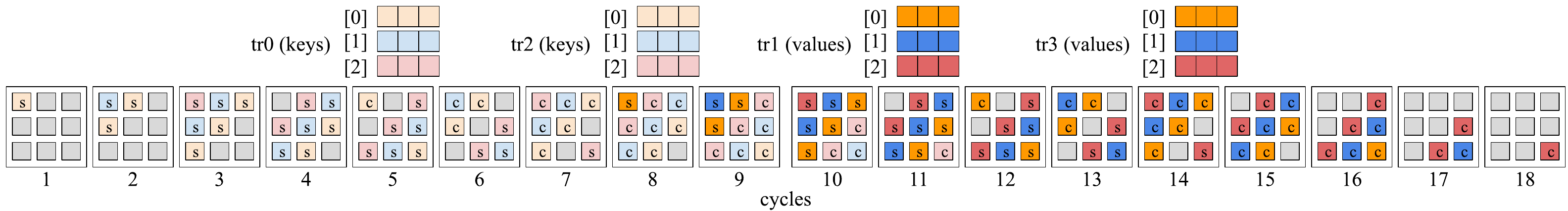}
  \caption{
    Cycle-by-Cycle Systolic Execution of Sorting Multiple Key-Value Lists in a 3$\times$3 Systolic Array --
    PEs performing key-value sorting are annotated with letter \TT{s}.
    PEs performing key-value compression are annotated with letter \TT{c}.
    PEs in gray color are idle.
    Otherwise, the color of a PE refers to a set of rows in matrix  registers
    that the PE is processing.
  }
  \label{fig:systolic-exec-all-rows}
  \vspace{-0.3cm}
\end{figure*}

%----------------
% Systolic execution of mssort*
%----------------

%\input{fig-mssortk-systolic-exec-example}
%\input{fig-mszipk-systolic-exec-example}

\subsection{Systolic Execution of Sorting Key-Value Chunks \\ in a Single Stream}
\label{sec-spz-uarch-mssort}

% explain input/output and components in the systolic array
Figure~\ref{fig:mssortk-systolic-exec} shows a 3$\times$3 systolic array
executing \TT{mssortk} instruction to sort two unsorted chunks of keys over
multiple cycles.
Initially, the two input chunks are located in the west and north sides of the
array.
Similar to a typical systolic execution of dense-dense GEMM, inputs are
staggered into the array.
Outputs come out from the east and south sides.
The dataflow through the array is the same as in dense-dense GEMM.
Each PE receives inputs from the west and north sides and sends outputs to the
east and south sides.

% explain how the data flow for mssortk
Input keys flow through the array in two passes: sorting and compressing.
In the sorting pass, keys in a chunk are sorted in an ascending order.
If a chunk contains duplicates, the sorting pass combines them into one valid
output key, and the rest become invalid outputs to be excluded.
The invalid outputs may exist in between valid outputs after the sorting pass,
so the compressing pass places valid outputs consecutively starting from the
first position of an output chunk and moves invalid outputs to the end.
For example, in Figure~\ref{fig:mssortk-systolic-exec}, after the sorting pass,
the north-side inputs \{5,~8,~5\} come out in the east side as \{5,~\TT{d},~8\}
(i.e., \TT{d} indicates excluded duplicated key(s)).
In the compressing pass, the partial output chunk is pushed into the array from
the west side, and the final output chunk \{5,~8,~\TT{d}\} comes out from the
south side.

% explain PE states
When sorting keys, we need to store the key reordering so that their values can
be shuffled to correct positions later.
In each PE, we encode and store the direction in which west-side and north-side
keys are routed towards and whether the keys are duplicate.
Given a pair of input keys, there are four possible states: (1) initial (no
data routing), (2) forwarding, (3) switching, and (4) combining.
The forwarding state encodes a PE routes west-side and north-side keys to the
east and south sides respectively.
The switching state encodes a PE routes west-side and north-side keys to the
south and east sides respectively.
The combining state encodes that two input keys are duplicate and that they are
combined into one valid key routed to the south side.
Each PE needs to store the states for both sorting and compressing passes.

% explain the algorithm
The west- and north-side input keys are sorted independently using the
bottom-left and top-right half of the systolic array.
PEs on the main diagonal are hard-coded to always switch inputs so that data
from two input chunks are not intermixed.
In other PEs, two input keys are compared.
The larger key is routed to the east, and the smaller key is routed to the
south (e.g., cycle 2 in Figure~\ref{fig:mssortk-systolic-exec}).
If keys are duplicate (e.g., cycle 5 in
Figure~\ref{fig:mssortk-systolic-exec}), a single combined valid key is sent to
the south port, and the east output is tagged as invalid key.
In subsequent PEs, the invalid key is considered larger than any valid key, so
it is always forwarded to the east.

% explain the counters
Instruction \TT{mssortk} updates special-purpose input and output counter
vector registers: \TT{W\_IC} (for the west input), \TT{N\_IC} (for the north
input), \TT{E\_OC} (for the east output) and \TT{S\_OC} (for the south output)
as keys come out from the array as shown in
Figure~\ref{fig:mssortk-systolic-exec} to count the number of processed input
keys and valid output keys.
Input counters (\TT{W\_IC} and \TT{N\_IC}) are updated in the sorting pass
while output counters (\TT{E\_OC} and \TT{S\_OC}) are updated in the
compressing pass.

% mssortv
Instruction \TT{mssortv} shuffles values based on a key reordering produced by
\TT{mssortk}.
Values are also passed through the array in two passes and directed based on
the states captured in each PE during the execution of \TT{mssortk}.
If the state is combining, two values are accumulated, and the accumulated
value is forwarded to a PE's south side.
Instruction \TT{mssortv} does not update the input and output counters.

%----------------
% Systolic execution of mszip*
%----------------
\subsection{Systolic Execution of Merging Key-Value Chunks \\ in a Single Stream}
\label{sec-spz-uarch-mszip}

% explain input and output
Figure~\ref{fig:mszipk-systolic-exec} shows the systolic execution of
\TT{mszipk} instruction in a 3$\times$3 array to merge two sorted key-value
chunks from a single stream.
Initially, the two input chunks are placed in the west and north sides.
Keys in the west side are ordered from bottom to top in an ascending order
while keys in the north side are ordered from left to right.
The final output chunk is stored in two parts.
For example, the part with smaller keys \{2,~3,~5\} is located in the east side
while the second part with larger keys \{8\} is stored in the south side.

% systolic data flow
Keys flow through the systolic array in two passes: merging and compressing.
The merging pass generates a merged list of sorted keys with invalid outputs
(i.e., caused by duplicate keys) potentially located in between valid output
keys.
The compressing pass then places valid output keys consecutively.
Unlike the sorting pass, the merging pass intermixes keys from both input
chunks, so PEs on the main diagonal work the same as other PEs instead of being
hard-coded to always switch inputs.
Similarly, larger keys flow to the east while smaller keys flow to the south
side.
The compressing pass works exactly the same as in the sorting operation.

% exclude keys that are not merged
Not all input keys can be merged into the output chunk.
Keys from an input chunk that are greater than all keys from the other chunk
need to be excluded since we do not know yet their positions in the output
stream as discussed in Section~\ref{sec-spz-isa-merge-spgemm}.
For example, in Figure~\ref{fig:mszipk-systolic-exec}, the key 9 in the west
input chunk is excluded from the output chunk (\TT{x} in cycle 4) since
it is greater than every key from the north-side chunk.
To detect keys to be excluded, each key is tagged with two extra bits to track
(1) from which input side the key comes from (source bit) and (2) whether the
key has been compared with another larger or equal key from the other input
chunk yet (merge bit).
The merge bit of a key is initially set to false and flipped to true when a
PE detects a larger or equal key from the other input side.
After the merging pass, if the merge bit is still false, the key is excluded
from the output chunk.

% explain the counters
Input and output counters are updated in the same way as in the sorting
operation.
\TT{W\_IC} and \TT{N\_IC} count the numbers of merged keys for the
west-side and north-side input chunks respectively.
\TT{E\_OC} and \TT{S\_OC} count the numbers of valid output keys in the
east-side and south-side output chunks respectively.
Instruction \TT{mszipv.tt} shuffles values based on the reordering of their
corresponding keys.

%----------------
% Multiple input lists
%----------------
\subsection{Merging and Sorting Key-Value Chunks \\ across Multiple Streams}
\label{sec-spz-uarch-multi-rows}

The execution of multiple micro-operations mapping to different streams can
overlap in SparseZipper's systolic array.
Figure~\ref{fig:systolic-exec-all-rows} shows the cycle-by-cycle systolic
execution of sorting key-value chunks from multiple streams.
Keys and values from multiple streams are mapped to different rows in matrix
registers.
Input keys and values from adjacent matrix register rows enter the systolic
array back-to-back in consecutive cycles since there is no data dependency
between micro-operations of an instruction.
There are one-cycle stalls in cycle 4 and cycle 11 since the systolic array
takes one extra cycle to route data from the west and south sides at the end of
a sorting pass to the east and north sides at the beginning of a compressing
pass.
The execution of \TT{mssortk}/\TT{mszipk} and \TT{mssortk}/\TT{mszipv}
instructions in a pair can overlap.
Since the latency of a micro-operation through the systolic array is fixed
(i.e., 2N+1 where N is the number of PEs in a row/column of the array), the
array can schedule to start the following \TT{mssortv}/\TT{mszipv} as soon as
the top-left-corner PE finishes its last key-compressing operation (e.g., in
cycle 8 in Figure~\ref{fig:systolic-exec-all-rows}).
However, the execution of different key-value instruction pairs processing
different input matrices of key-value data streams do not overlap to avoid
overwriting the array's output counters.
The counters must be read out to a vector register before the array can start
executing a new key-value instruction pair.

%----------------
% Systolic array details
%----------------
\subsection{Hardware Changes to the Baseline Systolic Array}
\label{sec-spz-uarch-ext}
\begin{figure*}[tp]
  \centering
  \includegraphics[width=\tw]{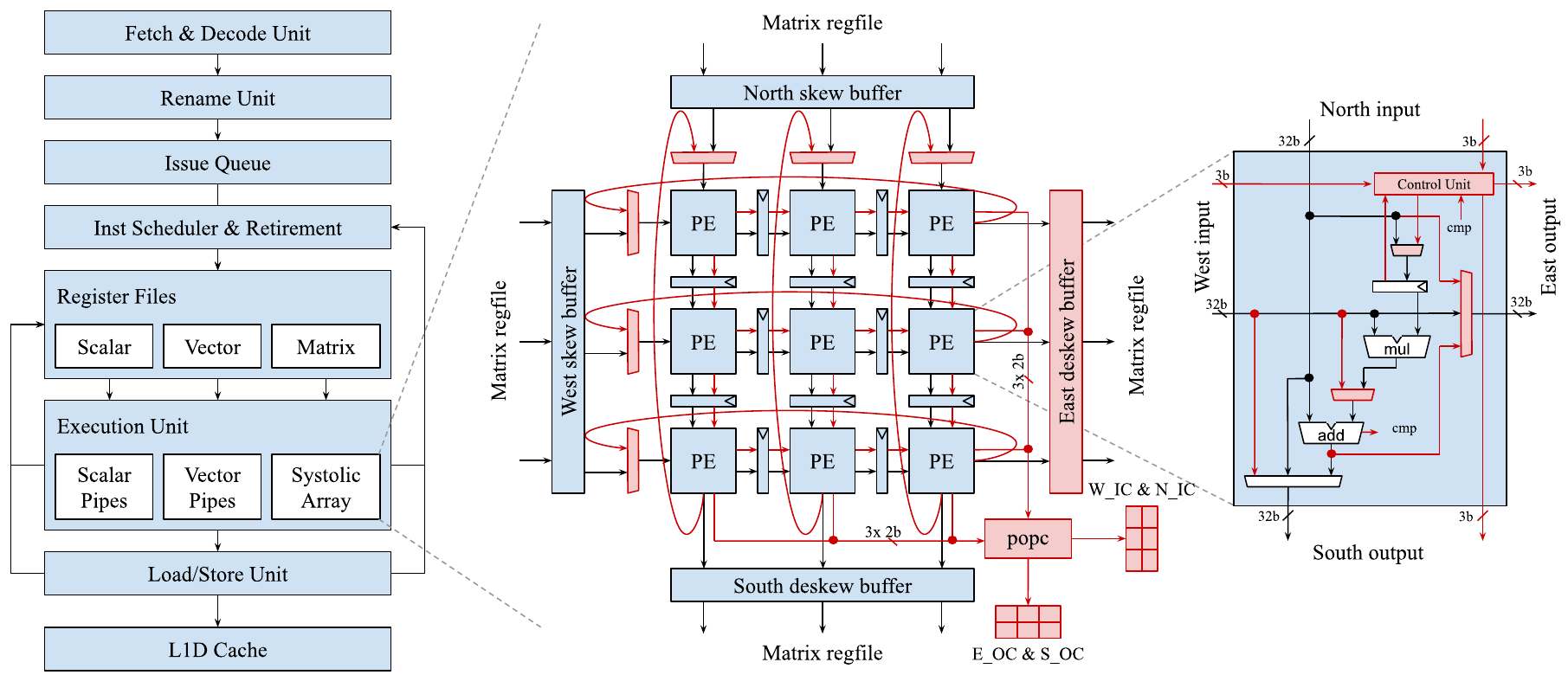}
  \caption[SparseZipper Systolic Array Micro-architecture]{
    SparseZipper Systolic Array Micro-architecture --
    Components and wires added to support sparse computation are in red;
    %PE~=~processing element;
    popc~=~population counting logic;
    W\_IC \& N\_IC~=~west \& north input counters;
    E\_OC \& S\_OC~=~east \& south output counters.
  }
  \label{fig:spz-sparse-zipper-uarch}
  \vspace{-0.2cm}
\end{figure*}

Figure~\ref{fig:spz-sparse-zipper-uarch} shows micro-architectural changes to
support SparseZipper in the baseline systolic array specialized for dense-dense
GEMM.
We add a second write port to the matrix register file as the sorting and
merging instructions have two output operands.
Since each physical matrix register is quite large (e.g., 1KB for a
16$\times$16 32-bit-element matrix register), the matrix register file may
consist of multiple SRAM banks, one for each physical matrix register.
Therefore, adding an additional write port to the register file simply requires
an extra crossbar instead of adding an extra write port to each SRAM bank which
would incur significant area overheads.
In order to retrieve the east-side output data from the systolic array, we add
a second deskew buffer.

% additional control signals and the loop-back paths
Additional control bits (i.e., source, duplicate, and merge bits) are tagged
along with data flowing through the systolic array as described in
Section~\ref{sec-spz-uarch-mszip}.
We add a three-bit control between any two PEs to the existing data paths.
For routing data between a sorting/merging pass and a compressing pass, two
loop-back paths are added to connect east and south output sides to the west
and north input sides respectively.
Each path is pipelined via an extra register to account for its long distance
between two sides of the systolic array.

% popc logic and input/output vectors of counters
Four input and output vectors of N counters are added to track the number of
valid input and output elements for N rows of matrix registers.
Each counter counts up to N (i.e., the number of elements in a row), so a
vector of N counters is $N \times \log_2 N$ bit wide.
The population counting logic uses output control signals from the systolic
array and increments corresponding input/output counters.

% inside each PE
In each PE, we slightly modify the existing adder to support comparison of
input keys.
An additional control unit uses the comparison outcome to make routing
decisions (i.e., forwarding, switching, and combining) and route data from
input to output ports by controlling the two output multiplexers.
The control unit also updates the duplicate and merge bits based on the source
bit and the comparison outcome.
We use the same adder for adding up values for \TT{mssortv} and \TT{mszipv}
instructions in case of combining inputs.
We repurpose the weight register in each PE to store routing states.
Each state requires two bits to encode.
Each pair of rows from two input matrix registers needs to store two states for
their sorting/merging and compressing passes.
Therefore, for N pairs of rows, we need a total of $N \times 4$ bits for all
the routing states (e.g., 64 bits for a hardware vector length of 16
elements).

%=========================================================================
% SparseZipper: Evaluation methodology
%=========================================================================

\section{Evaluation Methodology}
\label{sec-spz-methodology}

In this section, we describe our simulated systems, cycle-level modeling
methodology, and matrix datasets used to evaluate the performance of
SparseZipper.

\subsection{Simulated Systems}
%=========================================================================
% tbl-gem5.tex
%=========================================================================

\begin{table}[tp]
  \centering
  \cbxsetfontsize{8pt}
  \tabcolsep 3pt

  \caption[SparseZipper Baseline System Configuration]{
    Baseline System Configuration
  }
  % gem5 parameters
  \begin{tabular}{l>{\raggedright\arraybackslash}p{0.78\cw}}
    \toprule
    \BF{CPU}            & \textbullet~~RISC-V ISA (RV64GC) \newline
                          \textbullet~~8-way out-of-order issue \newline
                          \textbullet~~72-entry LQ, 56-entry SQ, 96-entry IQ \& 224-entry ROB \newline
                          \textbullet~~180 physical integer, 168 physical floating-point \& 128 physical 512-bit vector registers \newline
                          \textbullet~~Two 512-bit-wide SIMD execution units
                          \\\midrule
    \BF{Matrix Unit}    & \textbullet~~A systolic array with 16$\times$16 processing elements (PEs) \newline
                          \textbullet~~Each PE has a single-precision MAC unit \newline
                          \textbullet~~16 physical matrix registers
                          \\\midrule
    \BF{Caches}         & \textbullet~~L1I cache: 8-way, 32KB \& 2-cycle hit latency  \newline
                          \textbullet~~L1D cache: 8-way, 32KB \& 2-cycle hit latency  \newline
                          \textbullet~~L2 cache: 4-way, 4-bank, 256KB \& 8-cycle hit latency \newline
                          \textbullet~~LLC: 8-way, 8-bank, 512KB \& 8-cycle hit latency
                          \\\midrule
    \BF{Memory}         & DDR4-2400   \\
    \bottomrule
  \end{tabular}
  \vspace{0.1cm}

  \raggedright
  LQ~=~Load queue;
  SQ~=~Store queue;
  IQ~=~Issue queue;
  ROB~=~Reorder buffer;
  LLC~=~Last-level cache;
  MAC~=~multiply-accumulate unit.
  \label{tbl-spz-gem5}
  \vspace{-0.1cm}
\end{table}

\begin{table}[tp]
  \centering
  \cbxsetfontsize{8pt}
  \tabcolsep 3pt %2.2pt

  \caption[SparseZipper Evaluated Matrix Datasets]{
    Evaluated Datasets
  }

  \begin{tabular}{lrrrrrrr}
    \toprule
    % first row
    \multicolumn{1}{c}{\multirow{3}{*}{\BF{Matrix}}} &
    \multicolumn{1}{c}{\multirow{3}{*}{\BF{\# Rows}}} &
    \multicolumn{1}{c}{\multirow{3}{*}{\BF{NNZs}}} &
    \multicolumn{1}{c}{\multirow{3}{*}{\BF{Density}}} &
    \multicolumn{2}{c}{\BF{Per Row}} &
    \multicolumn{2}{c}{\BF{Per 16 Rows}} \\
    \cmidrule(lrrrr){5-6}  \cmidrule(lr){7-8}
    % second row
    \multicolumn{1}{c}{}                                &
    \multicolumn{1}{c}{}                                &
    \multicolumn{1}{c}{}                                &
    \multicolumn{1}{c}{}                                &
    \multicolumn{1}{c}{\BF{Avg}}                        &
    \multicolumn{1}{c}{\BF{Avg Out}}                    &
    \multicolumn{1}{c}{\BF{Avg}}                        &
    \multicolumn{1}{c}{\BF{Work}}                   \\
    %\multicolumn{1}{c}{\BF{Avg Work}}                   \\
    % third row
    \multicolumn{1}{c}{}                &
    \multicolumn{1}{c}{}                &
    \multicolumn{1}{c}{}                &
    \multicolumn{1}{c}{}                &
    \multicolumn{1}{c}{\BF{Work}}       &
    \multicolumn{1}{c}{\BF{NNZ}}        &
    \multicolumn{1}{c}{\BF{Work}}       &
    \multicolumn{1}{c}{\BF{Var}}  \\
    %\multicolumn{1}{c}{\BF{Variation}}  \\
    \midrule
    % dataset rows   nnzs    density    avg_work  avg_nnz_out  avg_work_16_rows  work_variation
    p2p       &  63K &  148K & 3.78E-05 &   8.60  &   8.59     &  0.14K          & 2.26           \\
    wiki      &   8K &  104K & 1.51E-03 & 547.52  & 220.70     &  8.76K          & 2.06           \\
    soc       &  76K &  509K & 8.84E-05 & 526.09  & 271.20     &  8.48K          & 1.43           \\
    ca-cm     &  23K &  187K & 3.49E-04 & 178.66  & 101.82     &  2.86K          & 1.35           \\
    ndwww     & 326K &  930K & 8.76E-06 &  29.42  &  12.63     &  0.78K          & 1.30           \\
    patents   & 241K &  561K & 9.69E-06 &  10.83  &   9.48     &  0.20K          & 1.29           \\
    ca-cs     & 227K & 1628K & 3.15E-05 & 164.38  &  72.68     &  2.63K          & 0.98           \\
    email     &  37K &  184K & 1.37E-04 & 163.04  &  89.30     &  2.64K          & 0.88           \\
    scircuit  & 171K &  959K & 3.28E-05 &  50.74  &  30.54     &  0.81K          & 0.48           \\
    bcsstk17  &  11K &  220K & 1.83E-03 & 445.71  &  56.58     &  7.13K          & 0.38           \\
    usroads   & 129K &  331K & 1.98E-05 &   7.18  &   5.45     &  0.11K          & 0.31           \\
    p3d       &  14K &  353K & 1.93E-03 & 870.85  & 218.85     & 13.93K          & 0.24           \\
    cage11    &  39K &  560K & 3.66E-04 & 225.13  &  97.59     &  3.60K          & 0.08           \\
    m133-b3   & 200K &  800K & 2.00E-05 &  16.00  &  15.90     &  0.26K          & 0.00           \\
    \bottomrule
  \end{tabular}

  \vspace{0.1cm}
  %p2p~=~p2p-Gnutella31;
  %wiki~=~wiki-Vote;
  %soc~=~soc-Epinions1;
  %ca-cm~=~ca-CondMat;
  %ndwww~=~NotreDame\_www;
  %patents~=~patents\_main;
  %ca-cs~=~coAuthorsCiteseer;
  %email~=~email-Enron;
  %p3d~=~poisson3Da;
  %Avg~=~average;
  \raggedright
  Var~=~coefficient variation, ratio of the standard deviation to the mean;
  Density~=~ratio of non-zero values to all values in a matrix;
  NNZ~=~number of non-zero values;
  Avg Out NNZ~=~average number of non-zero values in an output matrix row;
  Work~=~number of multiplications needed to compute one output row or one
  group of 16 consecutive output rows.
  %In this work, we multiply each square matrix with itself.
  %Table entries are sorted by the avg work variation in a descending order.
  %All matrices are square, so the number of rows is equal to the number of
  %columns.
  %For calculating the work, we assume multiplying each matrix with itself.
  %Work variation~=~coefficient variation, a ratio of the standard deviation
  %to the mean, of per-output-row work in a group of 16 consecutive rows.
  %The average work variation is the mean of all coefficient variations of all
  %16-output-row groups.
  
  \label{tab-spz-datasets}
  \vspace{-0.1cm}
\end{table}

We use gem5~\cite{binkert-gem5-sigarch2011, lowe-gem5-2020,
  ta-simulating-riscv-gem5-2018} to evaluate the performance of
SparseZipper. We use gem5's out-of-order core and configure it to model
an aggressive high-performance out-of-order Intel CPU with
state-of-the-art SIMD extensions (see Table~\ref{tbl-spz-gem5} for
baseline configuration). We model two 512-bit-wide SIMD execution units
integrated into the CPU pipeline with support for speculative
out-of-order execution of vector instructions. The simulated cache
subsystem is based on the Arm AMBA 5 CHI cache model provided in
gem5~\cite{chi-gem5}.

% for accelerating compute-intensive workloads like GEMM in high-end
% servers and data centers

% baseline matrix engine
\BF{Baseline systolic array for dense-dense GEMM --} As in previous
work~\cite{jeong-rasa-dac2021}, we model a systolic array with
16$\times$16 PEs similar in spirit to an implementation of Intel AMX in
Intel Saphhire Rapids~\cite{nassif-intel-sapphire-isscc2022}. Each PE
consists of a single-precision multiply-accumulate (MAC) unit with a
latency of four CPU cycles. There are 16 physical matrix registers, each
storing 16$\times$16 32-bit data.
% for a total of 1KB.
The baseline matrix register file supports two read ports and one write
port.
% Matrix registers can be renamed to avoid false dependencies.
% for matrix-matrix multiply instruction.
Since matrix registers are quite large, a reasonably area-efficient physical
implementation would include per-matrix-register 1r1w SRAMs and enough
crossbars for supporting two concurrent read and write accesses to two
different matrix registers.

% SparseZipper extension
%   - non-speculative instructions
%   - second write port
%   - indexed matrix load/store
\BF{Extended systolic array for SparseZipper --}
We model non-speculative execution of stream sorting and merging instructions
to simplify the hardware implementation.
% of special-purpose registers and systolic-array states.
These instructions wait until they are at the head of the ROB
% (i.e., no longer speculative)
before they are issued to the systolic array for execution. Once issued,
those instructions are placed into a retirement queue
% waiting for their
% execution (i.e., with no possible exception) to finish,
and subsequent instructions can continue to commit. We model extending
the matrix register file's crossbar to support the second write port. We
model a latency of one CPU cycle in each PE to process one pair of input
data when the PE executes the sorting and merging instructions since
those instructions do not use the PE's long-latency floating-point
multipler. Indexed matrix load and store instructions are broken into
row-wise micro-ops that are executed by the core's load-store unit.

\subsection{SpGEMM Implementations}

\BF{Scalar SpGEMM --} We evaluate two scalar row-wise implementations of
SpGEMM: \IT{scl-array} using dense
arrays~\cite{gilbert-sparse-matlab-1992} and \IT{scl-hash} using a hash
table with linear
probing~\cite{anh-hash-spgemm-2016,deveci-multithreaded-spgemm-2018} for
accumulating intermediate non-zero values in each output matrix row.
% In \IT{scl-hash}, we use with linear probing to solve hash collision.
% In both \IT{scl-array} and \IT{scl-hash},
After all intermediate non-zeros are accumulated for each output row,
they are sorted using a quick sort algorithm.

\BF{Vectorized Expand-Sort-Compress (ESC) SpGEMM --} We ported a
vectorized ESC implementation of SpGEMM, called \IT{vec-radix}, from
prior work~\cite{fevre-spgemm-rvv-2023}. The ESC algorithm was initially
proposed for performing SpGEMM on
GPUs~\cite{dalton-spgemm-gpu-2015,winter-adaptive-spgemm-gpu-ppopp2019}
and later adopted to vector
architectures~\cite{fevre-spgemm-rvv-2023,li-spgemm-vector-arch-mchpc2019}.
In ESC, multiple output rows are processed together to increase the
amount of parallelism, and the computation happens in three steps. First,
in the expansion step, results of multiplications are expanded in triples
of row index, column index, and value. Second, The list of triples are
sorted by their row and then column indices. This sorting step is often
vectorized using a fast radix sort~\cite{zaha-vector-rsort-sc1991}.
% which is vector-friendly.
Third, triples with duplicate keys (i.e., same row and column indices) are
compressed into one entry by accumulating the values.
In \IT{vec-radix}, there is a preprocessing step that calculates the amount of
work per block of output rows, determines the best block size, and allocates
enough temporary space for all intermediate results in a block.
Smaller block sizes limit the amount of parallelism,
% that can be vectorized
while larger block sizes can lead to thrashing the caches. We sweep the
block size for each input matrix and report the best performing
configuration.

\BF{Merge-based SpGEMM using SparseZipper --}
We implemented two versions of the merge-based row-wise dataflow SpGEMM using
the SparseZipper ISA extension: \IT{spz} and \IT{spz-rsort}.
In both versions, a preprocessing step calculates the amount of work for each
output row to allocate enough temporary memory space for intermediate results.
\IT{spz-rsort} additionally sorts row indices by the amount of
work calculated in the preprocessing step so that output rows with similar
amount of work can be processed together.
Only row indices are sorted, and the underlying matrix data is unchanged.
Once all output rows are computed, they are re-ordered by their row indices.
The sorting is done using a quick sort routine from the C++ standard library.
In both \IT{spz} and \IT{spz-rsort}, the expansion phase is vectorized using
the RISC-V vector extension while the merge phase is implemented using the
proposed SparseZipper instructions.

\subsection{Matrix Datasets}

We evaluate SparseZipper using matrices from
SuiteSparse~\cite{davis-suitesparse-2011} across multiple domains such as
road networks, scientific simulations, and social networks (see
Table~\ref{tab-spz-datasets}). This collection of matrices represents a
variety of sparsity levels and patterns. As in prior
work~\cite{srivastava-matraptor-micro2020, pal-outerspace-hpca2018,
  zhang-gamma-asplos2021}, we multiply each matrix with itself.
Table~\ref{tab-spz-datasets} reports the amount of work (i.e., the number
of multiplications needed) for each output row and for each group of 16
output rows. The table also shows the avarage number of non-zeros in
output matrices. The ratio of avarage work to the number of non-zeros per
row shows the degree in which duplicates in a stream of intermediate
non-zero values are compressed into a final stream of unique non-zero
values per output row.

%=========================================================================
% SparseZipper: Performance Evaluation
%=========================================================================

\section{Evaluation}
\label{sec-spz-perf-eval}

\begin{figure*}[tp]
    \centering
    \includegraphics[width=\tw]{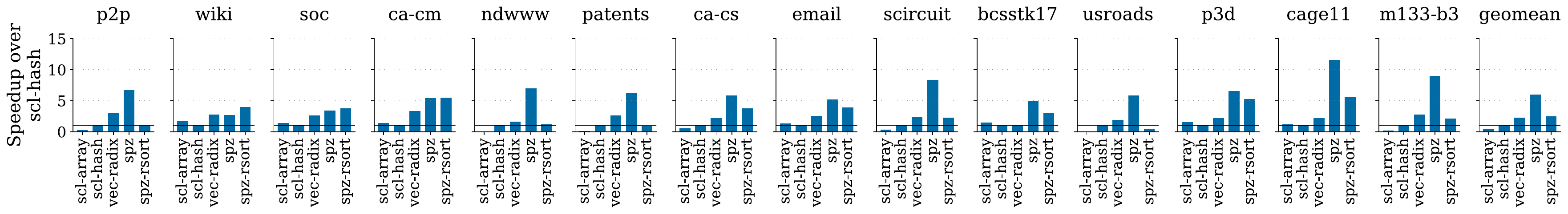}
    \caption[SparseZipper Performance]{
      Speedup over Scalar Baseline Using Hash Table
    }
    \label{fig-spz-result-speedup}
    \vspace{-0.1cm}
\end{figure*}

\begin{figure*}[tp]
    \centering
    \includegraphics[width=\tw]{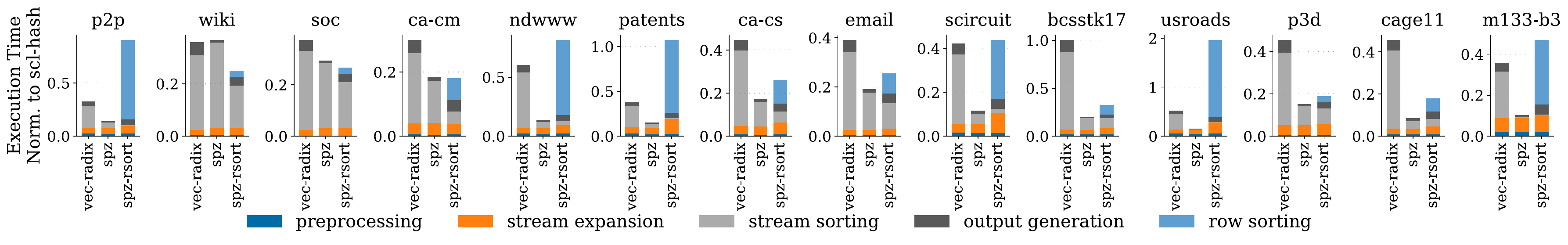}
    \caption[SparseZipper Execution Time Breakdown]{
      Execution Time Breakdown --
      The \IT{preprocessing} phase calculates per-row amount of work,
      divides the work into multiple row blocks (i.e., only in
      \TT{vec-radix}), and allocates memory space for intermediate results.
      The \IT{stream expansion} phase performs all multiplications and
      generates intermediate outputs.
      The \IT{stream sorting} phase sorts and compresses the intermediate
      outputs (i.e., only in \TT{spz-*}).
    }
    \label{fig-spz-result-exec-breakdown}
    \vspace{-0.1cm}
\end{figure*}

\begin{figure*}[tp]
    \centering
    \includegraphics[width=\tw]{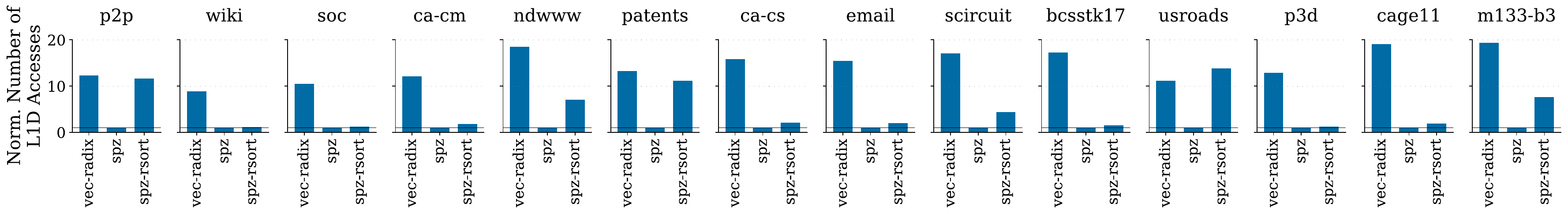}
    \caption[L1 Data Cache Accesses in SparseZipper]{
      Number of L1 Data Cache Accesses
    }
    \label{fig-spz-result-l1d-accs}
    \vspace{-0.3cm}
\end{figure*}

\begin{figure*}[tp]
    \centering
    \includegraphics[width=\tw]{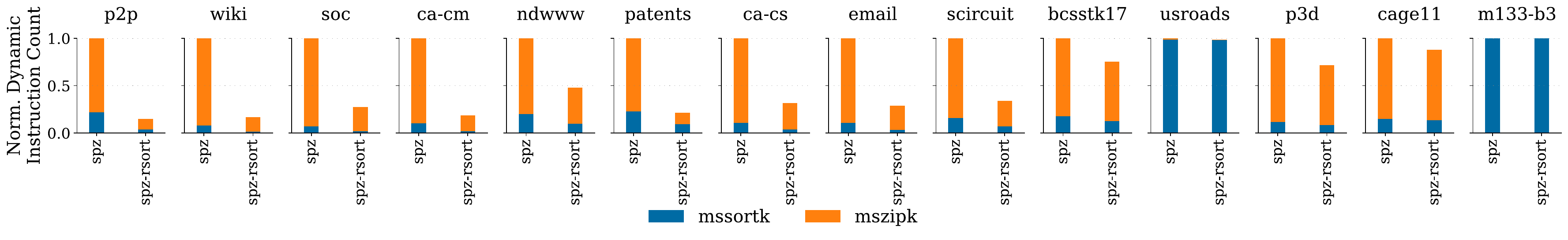}
    \caption[Dynamic Key Sorting and Merging Instruction Count in SparseZipper]{
      Number of Dynamic \TT{mssortk} and \TT{mszipk} Instructions
    }
    \label{fig-spz-result-mtx-inst-cnt}
    \vspace{-0.1cm}
\end{figure*}

In this section, we discuss cycle-level performance and first-order area
analyses of SparseZipper architecture.

\subsection{Performance Evaluation}

Figure~\ref{fig-spz-result-speedup} shows the relative performance of all
SpGEMM implementations evaluated in this work.
On average, \IT{spz} achieves 12.13$\times$, 5.98$\times$, 2.61$\times$
speedup over the three baseline versions \IT{scl-array}, \IT{scl-hash} and
\IT{vec-radix} respectively.

%----------
% scalar baseline
%----------
\BF{Scalar SpGEMM implementations --}
On average, \IT{scl-hash} is 2.03$\times$ faster than \IT{scl-array}.
For matrices that have relatively sparse outputs (e.g., \IT{p2p}, \IT{patents},
\IT{usroads}, and \IT{ndwww}), using a hash table to accumulate sparse non-zero
values is significantly more efficient than using dense arrays since each
output matrix row has a few non-zero values (i.e., shown in the average output
NNZ column in Table~\ref{tab-spz-datasets}).
In \IT{scl-array}, accesses to the dense array are scattered randomly, which
leads to low L1 data cache hit rates (e.g., less than 20\% for \IT{ndwww},
\IT{patents}, and \IT{usroads}).
In contrast, \IT{scl-hash} uses much smaller hash tables that help improve
significantly L1 data cache hit rates (e.g., close to 100\% for \IT{ndwww},
\IT{patents}, and \IT{usroads}).
A hash table's size is based on the amount of work per output matrix row
calculated in a preprocessing step.
For matrices that have relatively dense outputs (e.g., \IT{wiki}, \IT{soc},
\IT{bcsstk17}, and \IT{p3d}), \IT{scl-array} performs better than
\IT{scl-hash}.
The main reason is that accesses to a hash table for a relatively dense output
matrix cause frequent hash collisions that incur extra overheads.
In addition, those relatively dense matrices are typically smaller in sizes,
which helps improve the L1 cache hit rates.

%----------
% vector baseline
%----------
\BF{Vectorized SpGEMM implementation --}
On average, \IT{vec-radix} is 4.65$\times$ and 2.29$\times$ faster than
\IT{scl-array} and \IT{scl-hash} respectively.
Figure~\ref{fig-spz-result-exec-breakdown} shows the execution time breakdown
of \IT{vec-radix} in multiple steps.
Across all matrices, the combination of stream sorting and output generation,
which combines adjacent tuples with duplicate keys and generates final output
matrix rows, dominates the total execution time of \IT{vec-radix}.
For \IT{bcsstk17}, \IT{vec-radix} is slightly worse than \IT{scl-hash}.
The main reason is that \IT{bcsstk17} has a high ratio of per-row work to the
per-row number of output non-zeros, which indicates a high number of tuples
with duplicate keys finally compressed into a few non-zero values.
It is relatively inefficient to sort uncompressed key-value tuples with many
duplicate keys in the stream sorting step.

%----------
% matrix version
%----------
\BF{Merge-based SpGEMM using SparseZipper --}
% observation
The \IT{spz} version is 2.60$\times$ faster than the \IT{vec-radix}
implementation.
Figure~\ref{fig-spz-result-exec-breakdown} shows the execution time breakdown
of \IT{spz} in multiple steps.
The preprocessing and stream expansion steps in \IT{spz} are similar to
the ones in \IT{vec-radix}.
The proposed sorting and merging instructions targets to reduce the execution
time of the stream sorting step which dominates the execution time of
\IT{vec-radix}.
This reduction is shown in Figure~\ref{fig-spz-result-exec-breakdown} in almost
all matrices except \IT{wiki}.
It is important to note that the execution time for output generation is
decreased as well since \IT{spz} combines tuples with duplicate keys
while performing a merge sort on those tuples.
This avoids a separate compression step which is part of the output generation
in \IT{vec-radix}.

% why more efficient than vector baseline
One key reason for the higher performance is that \IT{spz} loads and stores
chunks of consecutive data using the proposed indexed matrix load-store
instructions.
Each row of a matrix register is loaded and stored using a unit-stride vector
memory micro-operation that minimizes the number of cache line accesses per
key-value chunk.
In contrast, \IT{vec-radix} uses a vectorized radix sort algorithm that
performs both long-stride and indexed vector memory accesses that span across
multiple cache lines, which results in multiple cache line accesses per vector
memory instruction.
Figure~\ref{fig-spz-result-l1d-accs} shows the significant reduction in the
number of L1 data cache accesses between \IT{vec-radix} and \IT{spz}
across all matrices.

%% explain the hit rate difference due to per-row work
%Despite efficient unit-stride vector memory accesses, \IT{spz} has lower
%L1 data cache hit rates than \IT{vec-radix} as shown in
%Figure~\ref{fig-spz-result-hit-rate} for some matrices (e.g., \IT{wiki},
%\IT{soc}, \IT{email}, \IT{bcsstk17}, and \IT{p3d}).
%Those matrices have relatively large numbers of intermediate results to merge
%per output matrix row, as shown in the per-row average work in
%Table~\ref{tab-spz-datasets}.
%Since \IT{spz} exploits parallelism across multiple output matrix rows
%(i.e., up to VLEN number of rows can be processed in parallel), \IT{spz}
%has larger aggregate working set (i.e., the average work per 16 rows in
%Table~\ref{tab-spz-datasets}) at a time than \IT{vec-radix}, which causes more
%L1 data cache misses.
%However, the latency of those misses can be hidden since indexed matrix load
%instructions can be issued out of order as soon as the next address offsets are
%produced by either \TT{mssortk} or \TT{mszipk} instruction.
%For other matrices (e.g., \IT{p2p}, \IT{patents}, \IT{usroads}, and
%\IT{m133-b3}) that have relatively low amount of work per row, the L1 data
%cache hit rate in \IT{spz} is close to the one in \IT{vec-radix}.

% explain some outliers: wiki and soc using the work variation
Since \IT{spz} processes a group of multiple streams in lock step, any
variation in lengths of those streams could impact its performance which is
determined by the processing time of the longest stream in the group.
Table~\ref{tab-spz-datasets} shows the work variation, a ratio of the work
standard deviation to the work mean, within a group of 16 consecutive matrix
rows.
The higher the work variation is, the more unbalanced the stream lengths of
adjacent matrix rows in a group are.
The relatively high work variation in \IT{wiki} and \IT{soc} explains the
relatively low performance of \IT{spz} compared to \IT{vec-radix}.
Although matrix \IT{p2p} has the highest work variation, \IT{spz}
performs well for this matrix since the average per-row work is low.
This low per-row amount of work minimizes the performance impact of high work
variation since it takes, on average, one iteration to finish processing one
key-value stream in \IT{p2p}.

% explain spz-rsort
To further demonstrate the performance impact of high work variation, we
sort matrix row indices by per-row amount of work in \IT{spz-rsort}.
It is important to note that we do not actually shuffle an input matrix's data
but simply sort row indices.
Rows with similar amount of work are then processed together.
At the end, it is necessary to shuffle the output matrix's data based on row
indices so that the final output data are sorted by their row indices.
Figure~\ref{fig-spz-result-exec-breakdown} shows the execution breakdown of
\IT{spz-rsort}.
By processing rows with similar amount of work together, the stream sorting
time in \IT{spz-rsort} is significantly reduced for matrices that have
high work variation (e.g., \IT{wiki}, \IT{soc}, \IT{ndww}, and \IT{ca-cmd}).
Figure~\ref{fig-spz-result-mtx-inst-cnt} shows the reduction in dynamic
instruction counts of \TT{mssortk} and \TT{mszipk} across matrices with high
work variation.
This reduction correlates to less number of iterations required to sort and
merge key-value streams due to more balanced work across rows in a group.
For \IT{cage11}, \IT{spz-rsort} results in a minimal reduction in the
stream sorting time since it has low work variation.
For \IT{usroads} and \IT{m133-b3}, since their average amount of work per row
is less than the vector length (i.e., 16), \IT{spz} and
\IT{spz-rsort} finish sorting each stream in one iteration on average
(i.e., only a few dynamic \TT{mszipk} instructions).

The row sorting and output data shuffling cause significant overheads in
\IT{spz-rsort}.
Row indices are sorted by a serial quick-sort routine provided in the standard
C++ library, which explains its high execution time.
Future work may extend SparseZipper to include instructions that are similar
to the stream merging and sorting for accelerating a standard merge-sort
routine that could potentially lower the row index sorting overhead.
In addition, processing rows in an order different from how their data are laid
out in memory causes a slight increase in the stream expansion time (e.g., in
\IT{patents} and \IT{scircuit}) due to poor spatial locality between rows.

\subsection{Area Evaluation}
\begin{table}[tp]
  \centering
  \cbxsetfontsize{8pt}
  \tabcolsep 3pt

  \caption{
    Post-synthesis Area Estimates of 16$\times$16 SparseZipper Systolic Array with 512-bit Datapath
  }

  \begin{tabular}{lrrr}
    \toprule
    % first row
    \multicolumn{1}{c}{\multirow{2}{*}{\BF{Component}}}   &
    \multicolumn{1}{c}{\BF{Area}}                         &
    \multicolumn{1}{c}{\multirow{2}{*}{\BF{Baseline}}}    &
    \multicolumn{1}{c}{\BF{Sparse}} \\
    % second row
    \multicolumn{1}{c}{}                  &
    \multicolumn{1}{c}{\BF{(k~$\um^2$)}}  &
    \multicolumn{1}{c}{}                  &
    \multicolumn{1}{c}{\BF{Zipper}}  \\
    \midrule
    Baseline PE (with a 32-bit MAC unit)      &  0.45  & $\times$~256   &               \\
    SparseZipper PE (with a 32-bit MAC unit)  &  0.51  &                & $\times$~256  \\
    Skew buffer (16-lane)                     &  3.16  & $\times$~2     & $\times$~2    \\
    Deskew buffer (16-lane)                   &  3.16  & $\times$~1     & $\times$~2    \\
    Matrix register (16 $\times$ 512b)        &  0.96  & $\times$~16    & $\times$~16   \\
    Popcount logic                                &  0.45  &                & $\times$~1    \\
    \midrule
    % last row
    \multicolumn{2}{l}{\BF{Total}}                              & 140.16 & 158.00       \\
    \multicolumn{2}{l}{\BF{SparseZipper vs. baseline overhead}} &        & \BF{12.72\%} \\
    \bottomrule
  \end{tabular}

    %PE~=~processing element;
    %MAC~=~multiply-accumulate;
    %16$\times$16 systolic array and 512-bit vector length.

  \label{tab-spz-area-result}
  \vspace{-0.3cm}
\end{table}

%    Baseline PE                         &  1.50  & $\times$~256   &               \\
%    SparseZipper PE                     &  1.61  &                & $\times$~256  \\
%    Skew buffer (16-lane)               &  6.32  & $\times$~2     & $\times$~2    \\
%    Deskew buffer (16-lane)             &  6.32  & $\times$~1     & $\times$~2    \\
%    Matrix register (16 $\times$ 512b)  &  0.96  & $\times$~16    & $\times$~16   \\
%    Popc logic                          &  0.45  &                & $\times$~1    \\
%    \midrule
%    % last row
%    \multicolumn{2}{l}{\BF{Total}}                              & 419.15 & 453.78       \\
%    \multicolumn{2}{l}{\BF{SparseZipper vs. baseline overhead}} &        & \BF{8.26\%}  \\

\BF{Methodology --}
We use a post-synthesis component-level area modeling methodology to evaluate
area overheads of hardware added to a baseline 16$\times$16 systolic array for
SparseZipper.
We implement area-significant components of the systolic array in RTL and
synthesize them using a 12nm standard-cell library.
Each PE includes a 32-bit single-precision floating-point MAC unit.
We model control logic added to a PE to support for the stream sorting and
merging operations.
Each skew/deskew buffer is used to stagger input and output data coming in and
going out of the systolic array.
We model each skew/deskew buffer as an array of 16 shift registers using
flip-flops with their sizes ranging from one to 16 entries.
We model 16 rows, each is 512-bit wide (16$\times$ 32-bit data elements), in a
SRAM-based matrix register and a total of 16 physical matrix registers.
Regarding the popcount logic, we implemented an array of 16 five-bit counters
(counting up to 16) and a list of counter vector registers (16 $\times$ 5 bits
per register).

\BF{Area overheads of SparseZipper --}
Table~\ref{tab-spz-area-result} shows the detailed area comparison between
SparseZipper and the baseline using our first-order component-based area
modeling methodology.
In overall, a 16$\times$16 SparseZipper implementation adds around 12.72\% area
overhead compared to the baseline implementation with the same systolic array's
dimensions.
When considering a complete system including an out-of-order core, its vector
engine, and its caches, we expect the percentage of extra area added to the
baseline systolic array for supporting SparseZipper to be much lower.

%=========================================================================
% SparseZipper: Related work
%=========================================================================

\section{Related Work}
\label{sec-spz-related-work}

\BF{Extending systolic arrays for sparse computation --}
NVIDIA introduced sparse tensor cores that accelerate sparse-dense GEMM with
one matrix having a specific 2:4 sparsity pattern (two zeros out of four
contiguous
elements)~\cite{pool-nvidia-sparse-tensor-2021,choquette-nvidia-tensor-core-2021}.
VEGETA supports more flexible sparsity structure (i.e., N:4 and N:M) on a
systolic array~\cite{jeong-vegeta-hpca2023}.
When performing sparse-dense GEMM, VEGETA keeps the sparse matrix stationary
and streams the dense matrix into a systolic array.
Each PE performs index matching to either skip or multiply two input values.
The output matrix is stored in the dense format.
Similar to SparseZipper, VEGETA is a fine-grain GEMM accelerator using a matrix
ISA extension.
Unlike NVIDIA sparse tensor cores and VEGETA, SparseZipper targets sparse-sparse
GEMM on highly sparse matrices with unstructured sparsity structures.
Processing such sparse-sparse GEMM using the sparse tensor core and VEGETA
would require several orders of magnitude more multiplications than using a
row-wise dataflow SpGEMM since matrices are highly sparse.
SparseZipper complements the sparse tensor core and VEGETA by enabling efficient execution of higher unstructured sparsity.

%Figure~\ref{fig-spz-num-mult} shows the computation demand (i.e., in the number
%of multiplications) of multiplying two matrices stored in different sparsity
%compression levels.
%SparseZipper stores both input matrices in a sparse format, which is the
%baseline in the figure.
%Sparse(1:4)-dense and sparse(1:16)-dense configurations represent two
%compression levels that can be supported in VEGETA.
%Dense-dense configuration represents a conventional systolic array that
%processes matrices in a dense format.
%Since our target matrix datasets are high sparse, the amount of computation
%needed to multiply a matrix stored in a sparse format with another matrix
%stored in a dense format is multiple orders of magnitude higher than the case
%in which both matrices are stored in a sparse format.
%SparseZipper complements the NVIDIA's sparse tensor core and VEGETA by
%expanding the spectrum of sparsity level and structure supports towards highly
%sparse and unstructured matrices.

Sparse-TPU~\cite{he-sparse-tpu-ics2020} proposed an offline column packing
algorithm that merges sparse columns in a matrix to minimize the number of
zeros mapped to a systolic array.
Sparse-TPU supports conditional execution to skip multiplications for values
that do not have matching indices.
However, Sparse-TPU targets only sparse-matrix dense-vector multiplication, not
sparse-sparse GEMM.
STA~\cite{liu-sta-cal2020} proposed a new block-sparse format that targets
matrices with an upper limit on the number of non-zeros in a block of elements.
In contrast, SparseZipper can support unstructured sparsity structures.

\BF{Software-only proposal to accelerate SpGEMM on a systolic array --}
Guo et al. proposed a software-only tiling optimization for DNN-specific SpGEMM
on a systolic array~\cite{guo-sw-only-spgemm-sc2020}.
Their pruning algorithm enforces a particular tile-wise sparsity pattern so
that dense tiles can be mapped directly to an underlying systolic array without
any hardware support.
However, this pruning algorithm is specific to DNN and only works for sparse
matrices generated from pruned DNN models.
In contrast, SparseZipper targets more general sparse matrices from various
domains (e.g., graph analytics) that may not have a particular sparsity
pattern.

\BF{Coarse-grain GEMM accelerators --}
Google TPU~\cite{jouppi-datacenter-isca2017,teich-google-tpu-v2-blog2018,jouppi-google-tpu-v2-v3-cacm2020,jouppi-tpu-v4-isca2023},
Eyeriss~\cite{chen-eyeriss-isscc2016} and Amazon AWS neuron
core~\cite{aws-neuron-core-2023} are some examples of coarse-grain dense-dense
GEMM accelerators that are highly inefficient when performing sparse-sparse
GEMM due to their lack of sparse format support and ability to skip ineffectual
multiplications.
Previous work has proposed various coarse-grain SpGEMM accelerators mainly
based on three different dataflows: inner product (e.g.,
SIGMA~\cite{qin-sigma-hpca2020} and Extensor~\cite{hegde-extensor-micro2019}),
outer product (e.g., OuterSparse~\cite{pal-outerspace-hpca2018} and
SpArch~\cite{zhang-sparch-hpca2020}), and row-wise product (e.g.,
MatRaptor~\cite{srivastava-matraptor-micro2020} and
Gamma~\cite{zhang-gamma-asplos2021}).
SparseZipper takes a more programmable approach that extends an existing matrix
ISA to support both dense-dense and sparse-sparse GEMM without adding
significant hardware area overhead.

\BF{Fine-grain GEMM accelerators --}
Intel AMX~\cite{intel-amx-web,nassif-intel-sapphire-isscc2022}, Arm
SME~\cite{arm-sme-web}, and RISC-V matrix extension
proposal~\cite{riscv-mtx-ext-proposal-web} are examples of matrix ISA
extensions for accelerating dense-dense GEMM.
RASA is an academic proposal for integrating a systolic array into an
out-of-order processor for dense-dense GEMM~\cite{jeong-rasa-dac2021}.
SparseZipper extends such ISAs and micro-architectures to
efficiently support SpGEMM.
SparseCore proposed an ISA extension for sparse tensor computation by
introducing stream registers and merging/intersecting
instructions~\cite{rao-sparsecore-asplos2022}.
%The merging instructions in SparseZipper were inspired by SparseCore.
Unlike SparseCore adding stream registers, SparseZipper leverages existing
matrix registers to store key-value streams.
%Instead of adding stream registers, SparseZipper
%leverages existing matrix registers designed to store key-value streams.
%SparseZipper does not add dedicated sparse processing units for merging and
%intersecting key-value streams.
Instead of adding dedicated sparse processing units for merging key-value
streams, SparseZipper minimally modifies an existing systolic array.

%=========================================================================
% SparseZipper: Conclusion
%=========================================================================

\section{Conclusion}
\label{sec-spz-conclusion}

This paper has demonstrated performance benefits of minimally extending a
matrix ISA and a systolic array micro-architecture originally designed for
dense GEMM to support SpGEMM.
The SparseZipper ISA introduces new stream sorting and merging instructions to
enable sorting and merging key-value streams representing sparse vectors in
SpGEMM computation.
SparseZipper leverages existing matrix registers to store parts of key-value
streams and minimally extends a systolic array to implement the stream sorting
and merging instructions.
Future research can explore opportunities to add instructions specialized for
certain sparsity structures (e.g., structured sparsity in deep learning
workloads) and to improve the utilization of a systolic array when it performs
the sorting or merging instructions.

%-------------------------------------------------------------------------
% Back Matter
%-------------------------------------------------------------------------

%\begin{acknowledgments}
%  This work was supported in part by ...
%\end{acknowledgments}

\bibliographystyle{paper-sparse-mtx-xcel}
\balance
\bibliography{paper-sparse-mtx-xcel}

\begin{thebibliography}{10}

\bibitem{anh-hash-spgemm-2016}
P.~N.~Q. Anh, R.~Fan, and Y.~Wen.
\newblock Balanced Hashing and Efficient GPU Sparse General Matrix-Matrix Multiplication.
\newblock {\em Int'l Symp. on Supercomputing (ICS)}, Jun 2016.

\bibitem{arm-sme-web}
Introducing the Scalable Matrix Extension for the Armv9-A Architecture.
\newblock Online Webpage, 2021 (accessed Apr 2023).

\bibitem{aws-neuron-core-2023}
NeuronCore-v2 Architecture.
\newblock AWS Neuron Technical Reference Manual, 2023.

\bibitem{azad-triangle-count-2015}
A.~Azad, A.~Bulu{\c{c}}, and J.~Gilbert.
\newblock Parallel Triangle Counting and Enumeration Using Matrix Algebra.
\newblock pages 804--811, 2015.

\bibitem{binkert-gem5-sigarch2011}
N.~Binkert, B.~M. Beckmann, G.~Black, S.~K. Reinhardt, A.~Saidi, A.~Basu, J.~Hestness, D.~R. Hower, T.~Krishna, S.~Sardashti, R.~Sen, K.~Sewell, M.~Shoaib, N.~Vaish, M.~D. Hill, and D.~A. Wood.
\newblock The gem5 Simulator.
\newblock {\em SIGARCH Computer Architecture News (CAN)}, 39(2):1--7, Aug 2011.

\bibitem{canning-sparse-sim-1996}
A.~Canning, G.~Galli, F.~Mauri, A.~De~Vita, and R.~Car.
\newblock O(N) Tight-binding Molecular Dynamics on Massively Parallel Computers: an Orbital Decomposition Approach.
\newblock {\em Computer Physics Communications}, 94(2-3):89--102, 1996.

\bibitem{chen-eyeriss-isscc2016}
Y.-H. Chen, T.~Krishna, J.~Emer, and V.~Sze.
\newblock Eyeriss: An Energy-Efficient Reconfigurable Accelerator for Deep Convolutional Neural Networks.
\newblock {\em Int'l Solid-State Circuits Conf. (ISSCC)}, Feb 2016.

\bibitem{chen-eyeriss-v2-jetcas2019}
Y.-H. Chen, T.-J. Yang, J.~Emer, and V.~Sze.
\newblock Eyeriss v2: A Flexible Accelerator for Emerging Deep Neural Networks on Mobile Devices.
\newblock {\em IEEE Journal on Emerging and Selected Topics in Circuits and Systems (JETCAS)}, 9:292--308, Jun 2019.

\bibitem{choquette-tensor-core-nvidia-ieeemicro2021}
J.~Choquette, W.~Gandhi, O.~Giroux, N.~Stam, and R.~Krashinsky.
\newblock Nvidia A100 Tensor Core GPU: Performance and Innovation.
\newblock {\em IEEE Micro}, 41(2):29--35, 2021.

\bibitem{choquette-nvidia-tensor-core-2021}
J.~Choquette, W.~Gandhi, O.~Giroux, N.~Stam, and R.~Krashinsky.
\newblock Nvidia a100 Tensor Core GPU: Performance and Innovation.
\newblock {\em IEEE Micro}, 41(2):29--35, 2021.

\bibitem{dalberto-all-pair-spgemm-2007}
P.~D'alberto and A.~Nicolau.
\newblock R-Kleene: A High-Performance Divide-and-Conquer Algorithm for the All-Pair Shortest Path for Densely Connected Networks.
\newblock {\em Algorithmica}, 47:203--213, 2007.

\bibitem{dalton-spgemm-gpu-2015}
S.~Dalton, L.~Olson, and N.~Bell.
\newblock Optimizing Sparse Matrix—Matrix Multiplication for the GPU.
\newblock {\em ACM Trans. on Mathematical Software (TOMS)}, 41(4):1--20, 2015.

\bibitem{davis-graphblas-tmos2019}
T.~A. Davis.
\newblock Algorithm 1000: SuiteSparse: GraphBLAS: Graph Algorithms in the Language of Sparse Linear Algebra.
\newblock {\em ACM Trans. Math. Softw.}, 45(4), 2019.

\bibitem{davis-suitesparse-2011}
T.~A. Davis and Y.~Hu.
\newblock The University of Florida Sparse Matrix Collection.
\newblock {\em ACM Trans. Math. Softw.}, 38(1):1:1--1:25, Dec 2011.

\bibitem{deveci-multithreaded-spgemm-2018}
M.~Deveci, C.~Trott, and S.~Rajamanickam.
\newblock Multithreaded Sparse Matrix-Matrix Multiplication for Many-Core and GPU Architectures.
\newblock {\em Parallel Computing}, 78:33--46, 2018.

\bibitem{fevre-spgemm-rvv-2023}
V.~L. F{\`e}vre and M.~Casas.
\newblock Optimization of SpGEMM with RISCV-V Vector Instructions.
\newblock {\em Computing Research Repository (CoRR)}, Mar 2023.

\bibitem{galli-quantum-sim-1996}
G.~Galli.
\newblock Linear Scaling Methods for Electronic Structure Calculations and Quantum Molecular Dynamics Simulations.
\newblock {\em Current Opinion in Solid State and Materials Science}, 1(6):864--874, 1996.

\bibitem{chi-gem5}
Arm's AMBA 5 CHI Ruby Model in gem5.
\newblock Online Webpage, accessed Nov 20, 2021.

\bibitem{gilbert-sparse-matlab-1992}
J.~R. Gilbert, C.~Moler, and R.~Schreiber.
\newblock Sparse Matrices in MATLAB: Design and Implementation.
\newblock {\em SIAM Journal on Matrix Analysis and Applications}, 13(1):333--356, 1992.

\bibitem{guo-sw-only-spgemm-sc2020}
C.~Guo, B.~Y. Hsueh, J.~Leng, Y.~Qiu, Y.~Guan, Z.~Wang, X.~Jia, X.~Li, M.~Guo, and Y.~Zhu.
\newblock Accelerating Sparse DNN Models Without Hardware-Support via Tile-Wise Sparsity.
\newblock {\em Int'l Conf. on High Performance Networking and Computing (Supercomputing)}, Nov 2020.

\bibitem{han-deep-compress-arxiv2015}
S.~Han, H.~Mao, and W.~J. Dally.
\newblock Deep Compression: Compressing Deep Neural Networks with Pruning, Trained Quantization and Huffman Coding.
\newblock {\em Computing Research Repository (CoRR)}, Oct 2015.

\bibitem{he-sparse-tpu-ics2020}
X.~He, S.~Pal, A.~Amarnath, S.~Feng, D.-H. Park, A.~Rovinski, H.~Ye, Y.~Chen, R.~Dreslinski, and T.~Mudge.
\newblock Sparse-TPU: Adapting Systolic Arrays for Sparse Matrices.
\newblock {\em Int'l Symp. on Supercomputing (ICS)}, Jun 2020.

\bibitem{hegde-extensor-micro2019}
K.~Hegde, H.~Asghari-Moghaddam, M.~Pellauer, N.~Crago, A.~Jaleel, E.~Solomonik, J.~Emer, and C.~W. Fletcher.
\newblock Extensor: An accelerator for sparse tensor algebra.
\newblock {\em Int'l Symp. on Microarchitecture (MICRO)}, Oct 2019.

\bibitem{hoefler2011generic}
T.~Hoefler and M.~Snir.
\newblock Generic Topology Mapping Strategies for Large-scale Parallel Architectures.
\newblock {\em Int'l Symp. on Supercomputing (ICS)}, May 2011.

\bibitem{ibm-mmx-assist-web}
Matrix-Multiply Assist Best Practices Guide.
\newblock Online Webpage, 2021 (accessed Apr 2023).

\bibitem{intel-amx-web}
Intel Advanced Matrix Extensions Overview.
\newblock Online Webpage, (accessed Apr 2023).

\bibitem{itoh-order-n-spgemm-1995}
S.~Itoh, P.~Ordej{\'o}n, and R.~M. Martin.
\newblock Order-N Tight-binding Molecular Dynamics on Parallel Computers.
\newblock {\em Computer Physics Communications}, 88(2-3):173--185, 1995.

\bibitem{jeong-vegeta-hpca2023}
G.~Jeong, S.~Damani, A.~R. Bambhaniya, E.~Qin, C.~J. Hughes, S.~Subramoney, H.~Kim, and T.~Krishna.
\newblock VEGETA: Vertically-Integrated Extensions for Sparse/Dense GEMM Tile Acceleration on CPUs.
\newblock {\em Int'l Symp. on High-Performance Computer Architecture (HPCA)}, Mar 2023.

\bibitem{jeong-rasa-dac2021}
G.~Jeong, E.~Qin, A.~Samajdar, C.~J. Hughes, S.~Subramoney, H.~Kim, and T.~Krishna.
\newblock RASA: Efficient Register-Aware Systolic Array Matrix Engine for CPU.
\newblock {\em Design Automation Conf. (DAC)}, Nov 2021.

\bibitem{jouppi-tpu-v4-isca2023}
N.~Jouppi, G.~Kurian, S.~Li, P.~Ma, R.~Nagarajan, L.~Nai, N.~Patil, S.~Subramanian, A.~Swing, B.~Towles, et~al.
\newblock TPU v4: An Optically Reconfigurable Supercomputer for Machine Learning with Hardware Support for Embeddings.
\newblock {\em Int'l Symp. on Computer Architecture (ISCA)}, Jun 2023.

\bibitem{jouppi-google-tpu-v2-v3-cacm2020}
N.~P. Jouppi, D.~H. Yoon, G.~Kurian, S.~Li, N.~Patil, J.~Laudon, C.~Young, and D.~Patterson.
\newblock A Domain-Specific Supercomputer for Training Deep Neural Networks.
\newblock {\em Communications of the ACM}, 63(7):67--78, Jul 2020.

\bibitem{jouppi-datacenter-isca2017}
N.~P. Jouppi, C.~Young, N.~Patil, D.~Patterson, G.~Agrawal, R.~Bajwa, S.~Bates, S.~Bhatia, N.~Boden, A.~Borchers, et~al.
\newblock In-datacenter Performance Analysis of a Tensor Processing Unit.
\newblock {\em Int'l Symp. on Computer Architecture (ISCA)}, Jun 2017.

\bibitem{karypis-interior-point-alg-1994}
G.~Karypis, A.~Gupta, and V.~Kumar.
\newblock A Parallel Formulation of Interior Point Algorithms.
\newblock pages 204--213, 1994.

\bibitem{li-spgemm-vector-arch-mchpc2019}
J.~Li, F.~Wang, T.~Araki, and J.~Qiu.
\newblock Generalized Sparse Matrix-matrix Multiplication for Vector Engines and Graph Applications.
\newblock {\em IEEE/ACM Workshop on Memory Centric High Performance Computing (MCHPC)}, Nov 2019.

\bibitem{liu-sta-cal2020}
Z.-G. Liu, P.~N. Whatmough, and M.~Mattina.
\newblock Systolic Tensor Array: An Efficient Structured-Sparse GEMM Accelerator for Mobile CNN Inference.
\newblock {\em Computer Architecture Letters (CAL)}, 19(1):34--37, 2020.

\bibitem{lowe-gem5-2020}
J.~Lowe-Power, A.~M. Ahmad, A.~Akram, M.~Alian, R.~Amslinger, M.~Andreozzi, A.~Armejach, N.~Asmussen, B.~Beckmann, S.~Bharadwaj, G.~Black, G.~Bloom, B.~R. Bruce, D.~R. Carvalho, J.~Castrillon, L.~Chen, N.~Derumigny, S.~Diestelhorst, W.~Elsasser, C.~Escuin, M.~Fariborz, A.~Farmahini-Farahani, P.~Fotouhi, R.~Gambord, J.~Gandhi, D.~Gope, T.~Grass, A.~Gutierrez, B.~Hanindhito, A.~Hansson, S.~Haria, A.~Harris, T.~Hayes, A.~Herrera, M.~Horsnell, S.~A.~R. Jafri, R.~Jagtap, H.~Jang, R.~Jeyapaul, T.~M. Jones, M.~Jung, S.~Kannoth, H.~Khaleghzadeh, Y.~Kodama, T.~Krishna, T.~Marinelli, C.~Menard, A.~Mondelli, M.~Moreto, T.~Mück, O.~Naji, K.~Nathella, H.~Nguyen, N.~Nikoleris, L.~E. Olson, M.~Orr, B.~Pham, P.~Prieto, T.~Reddy, A.~Roelke, M.~Samani, A.~Sandberg, J.~Setoain, B.~Shingarov, M.~D. Sinclair, T.~Ta, R.~Thakur, G.~Travaglini, M.~Upton, N.~Vaish, I.~Vougioukas, W.~Wang, Z.~Wang, N.~Wehn, C.~Weis, D.~A. Wood, H.~Yoon, and Éder F.~Zulian.
\newblock The gem5 Simulator: Version 20.0+.
\newblock {\em arXiv preprint arXiv:2007.03152}, 2020.

\bibitem{nassif-intel-sapphire-isscc2022}
N.~Nassif, A.~O. Munch, C.~L. Molnar, G.~Pasdast, S.~V. Lyer, Z.~Yang, O.~Mendoza, M.~Huddart, S.~Venkataraman, S.~Kandula, et~al.
\newblock Sapphire Rapids: The Next-generation Intel Xeon Scalable Processor.
\newblock {\em Int'l Solid-State Circuits Conf. (ISSCC)}, Feb 2022.

\bibitem{naumov-dnn-model-arxiv2019}
M.~Naumov, D.~Mudigere, H.-J.~M. Shi, J.~Huang, N.~Sundaraman, J.~Park, X.~Wang, U.~Gupta, C.-J. Wu, A.~G. Azzolini, et~al.
\newblock Deep Learning Recommendation Model for Personalization and Recommendation Systems.
\newblock {\em Computing Research Repository (CoRR)}, May 2019.

\bibitem{pal-outerspace-hpca2018}
S.~Pal, J.~Beaumont, D.-H. Park, A.~Amarnath, S.~Feng, C.~Chakrabarti, H.-S. Kim, D.~Blaauw, T.~Mudge, and R.~Dreslinski.
\newblock OuterSPACE: An Outer Product Based Sparse Matrix Multiplication Accelerator.
\newblock {\em Int'l Symp. on High-Performance Computer Architecture (HPCA)}, Feb 2018.

\bibitem{penn-context-free-grammar-2006}
G.~Penn.
\newblock Efficient Transitive Closure of Sparse Matrices over Closed Semirings.
\newblock {\em Theoretical Computer Science}, 354(1):72--81, 2006.

\bibitem{pool-nvidia-sparse-tensor-2021}
J.~Pool, A.~Sawarkar, and J.~Rodge.
\newblock Accelerating Inference with Sparsity Using the NVIDIA Ampere Architecture and NVIDIA TensorRT.
\newblock NVIDIA Technical Report, 2021.

\bibitem{qin-sigma-hpca2020}
E.~Qin, A.~Samajdar, H.~Kwon, V.~Nadella, S.~Srinivasan, D.~Das, B.~Kaul, and T.~Krishna.
\newblock SIGMA: A Sparse and Irregular GEMM Accelerator with Flexible Interconnects for DNN Training.
\newblock {\em Int'l Symp. on High-Performance Computer Architecture (HPCA)}, Feb 2020.

\bibitem{rabin-maximum-matching-alg-1989}
M.~O. Rabin and V.~V. Vazirani.
\newblock Maximum Matchings in General Graphs through Randomization.
\newblock {\em Journal of algorithms}, 10(4):557--567, 1989.

\bibitem{rao-sparsecore-asplos2022}
G.~Rao, J.~Chen, J.~Yik, and X.~Qian.
\newblock SparseCore: Stream ISA and Processor Specialization for Sparse Computation.
\newblock {\em Int'l Conf. on Architectural Support for Programming Languages and Operating Systems (ASPLOS)}, Feb 2022.

\bibitem{reddi-mlperf-isca2020}
V.~J. Reddi, C.~Cheng, D.~Kanter, P.~Mattson, G.~Schmuelling, C.-J. Wu, B.~Anderson, M.~Breughe, M.~Charlebois, W.~Chou, et~al.
\newblock MLPerf Inference Benchmark.
\newblock {\em Int'l Symp. on Computer Architecture (ISCA)}, Jun 2020.

\bibitem{riscv-vector-ext}
RISC-V Vector Extension (Version 0.10).
\newblock Online Webpage, 2021.

\bibitem{riscv-mtx-ext-proposal-web}
RISC-V Matrix Extension Specification Proposal.
\newblock Online Webpage, (accessed Apr 2023).

\bibitem{shah-graph-spgemm-2007}
V.~B. Shah.
\newblock {\em An Interactive System for Combinatorial Scientific Computing with an Emphasis on Programmer Productivity}.
\newblock University of California, Santa Barbara, 2007.

\bibitem{shun-multicore-tc-2015}
J.~Shun and K.~Tangwongsan.
\newblock Multicore triangle computations without tuning.
\newblock {\em Int'l Conf. on Data Engineering (ICDE)}, Apr 2015.

\bibitem{srivastava-matraptor-micro2020}
N.~Srivastava, H.~Jin, J.~Liu, D.~Albonesi, and Z.~Zhang.
\newblock MatRaptor: A Sparse-Sparse Matrix Multiplication Accelerator Based on Row-Wise Product.
\newblock {\em Int'l Symp. on Microarchitecture (MICRO)}, Oct 2020.

\bibitem{ta-simulating-riscv-gem5-2018}
T.~Ta, L.~Cheng, and C.~Batten.
\newblock Simulating Multi-Core RISC-V Systems in gem5.
\newblock {\em Workshop on Computer Architecture Research with RISC-V}, 2018.

\bibitem{teich-google-tpu-v2-blog2018}
P.~Tiech.
\newblock Under the Hood of Google's TPU2 Machine Learning Clusters.
\newblock Online Article, May 2017 (accessed Feb 2020).

\bibitem{winter-adaptive-spgemm-gpu-ppopp2019}
M.~Winter, D.~Mlakar, R.~Zayer, H.-P. Seidel, and M.~Steinberger.
\newblock Adaptive Sparse Matrix-matrix Multiplication on the GPU.
\newblock {\em Symp. on Principles and practice of Parallel Programming (PPoPP)}, Feb 2019.

\bibitem{wu-ml-facebook-hpca2019}
C.-J. Wu, D.~Brooks, K.~Chen, D.~Chen, S.~Choudhury, M.~Dukhan, K.~Hazelwood, E.~Isaac, Y.~Jia, B.~Jia, et~al.
\newblock Machine Learning at Facebook: Understanding Inference at the Edge.
\newblock {\em Int'l Symp. on High-Performance Computer Architecture (HPCA)}, Mar 2019.

\bibitem{yamazaki-spgemm-schur-2010}
I.~Yamazaki and X.~S. Li.
\newblock On Techniques to Improve Robustness and Scalability of a Parallel Hybrid Linear Solver.
\newblock pages 421--434, 2010.

\bibitem{zaha-vector-rsort-sc1991}
M.~Zagha and G.~E. Blelloch.
\newblock Radix Sort for Vector Multiprocessors.
\newblock {\em Int'l Conf. on High Performance Networking and Computing (Supercomputing)}, 1991.

\bibitem{zhang-gamma-asplos2021}
G.~Zhang, N.~Attaluri, J.~S. Emer, and D.~Sanchez.
\newblock Gamma: Leveraging Gustavson’s Algorithm to Accelerate Sparse Matrix Multiplication.
\newblock {\em Int'l Conf. on Architectural Support for Programming Languages and Operating Systems (ASPLOS)}, Apr 2021.

\bibitem{zhang-sparch-hpca2020}
Z.~Zhang, H.~Wang, S.~Han, and W.~J. Dally.
\newblock SpArch: Efficient Architecture for Sparse Matrix Multiplication.
\newblock {\em Int'l Symp. on High-Performance Computer Architecture (HPCA)}, Feb 2020.

\end{thebibliography}

\end{document}